\documentclass[twocolumn]{aastex63}

\usepackage{savesym}
\usepackage{natbib}
\usepackage{amsmath,amssymb,amsthm,mathrsfs,dsfont}
\usepackage{graphicx}
\usepackage{subfigure}
\usepackage{captcont}
\usepackage{float}
\usepackage{booktabs}
\usepackage{epstopdf}
\newcommand{\yr}{{\rm \,yr}}

\newcommand{\gyr}{{\rm \,Gyr}}
\newcommand{\pyr}{{\rm yr^{-1}}}
\newcommand{\pgal}{{\rm gal^{-1}}}
\newcommand{\pmpc}{{\rm Mpc^{-3}}}

\newcommand{\msun}{{\rm \,M_{\odot}}}

\newcommand{\gal}{_{\rm gal}}
\newcommand{\bgal}{{_{\rm *}}}

\newcommand{\bh}{_{\rm BH}}
\newcommand{\gm}{_{\mathscr{F}}}

\newcommand{\lc}{_{\rm lc}}
\newcommand{\Lc}{^{\rm lc}}
\newcommand{\lr}{_{\rm lr}}

\newcommand{\Drain}{^{\rm drain}}
\newcommand{\enhance}{_{\rm enhance}}

\newcommand{\calE}{\mathcal{E}}

\newcommand{\rmc}{_{\rm c}}

\newcommand{\rmr}{_{\rm r}}
\newcommand{\rms}{_{\rm s}}
\newcommand{\rmt}{_{\rm t}}
\newcommand{\ndot}{{\mathscr{F}}}
\newcommand{\Glc}{{\mathscr{F}}^{\rm lc}}

\newcommand{\Gtot}{{\mathscr{F}}^{\rm consp}}
\newcommand{\GVinte}{{\mathbb{F}}_{\rm vol}}
\newcommand{\GVlcinte}{{\mathbb{F}}_{\rm vol}^{\rm lc}}
\newcommand{\GVtotinte}{{\mathbb{F}}_{\rm vol}^{\rm consp}}
\newcommand{\GVdiff}{d{\mathbb{F}}_{\rm vol}/d\log M_{\rm BH}}
\newcommand{\GlcVdiff}{d{\mathbb{F}}_{\rm vol}^{\rm lc}/d\log M_{\rm BH}}
\newcommand{\GtotVdiff}{d{\mathbb{F}}_{\rm vol}^{\rm consp}/d\log M_{\rm BH}}

\newcommand{\rmswl}{_{\rm swl}}
\newcommand{\rmpeak}{_{\rm peak}}

\newcommand{\Gfuel}{{\mathscr{F}}^{\rm consp}}
\newcommand{\GVfuelinte}{{\mathbb{F}}_{\rm vol}^{\rm consp}}
\newcommand{\GfuelVdiff}{d{\mathbb{F}}_{\rm vol}^{\rm consp}/d\log M_{\rm BH}}
\newcommand{\Glcmean}{{\langle \Glc \rangle}}
\newcommand{\Gfuelmean}{{\langle \Gfuel \rangle}}

\newcommand{\Lr}{^{\rm lr}}
\newcommand{\lw}{^{\rm lw}}
\newcommand{\full}{_{\rm full}}
\newcommand{\consp}{^{\rm consp}}
\newcommand{\tri}{_{\rm tri}}

\begin{document}

\title{Cosmic distributions of stellar tidal disruptions by massive black
holes at galactic centers}
\shorttitle{Cosmic distributions of stellar tidal disruption events}
\shortauthors{Chen, Yu, \& Lu}

\author{Yunfeng Chen}
\affiliation{Kavli Institute for Astronomy and Astrophysics, and School of Physics, Peking University, Beijing, 100871, China}

\author{Qingjuan Yu}
\affiliation{Kavli Institute for Astronomy and Astrophysics, and School of Physics, Peking University, Beijing, 100871, China}

\author{Youjun Lu}
\affiliation{National Astronomical Observatories, Chinese Academy of Sciences, Beijing, 100012, China}
\affiliation{School of Astronomy and Space Science, University of Chinese
Academy of Sciences, Beijing 100049, China}

\correspondingauthor{Qingjuan Yu}
\email{yuqj@pku.edu.cn}

\begin{abstract}
Stars can be consumed (either tidally disrupted or swallowed whole) by massive
black holes (MBHs) at galactic centers when they move into the vicinity of the
MBHs. In this study, we investigate the rates of stellar consumptions by
central MBHs and their cosmic distributions, including the effects of 
triaxial galaxy shapes in enhancing the reservoir of low-angular-momentum stars and
incorporating realistic galaxy distributions.  We find that the enhancement in
the stellar consumption rates due to triaxial galaxy shapes can be significant,
by a factor of $\sim3$ for MBH mass $M\bh\sim10^5$--$10^6\msun$ and up to more
than one order of magnitude for $M\bh\ga10^8\msun$. Only for
$M\bh\la10^7\msun$ are the stellar consumption rates significantly higher in
galaxies with steeper inner surface brightness profiles. The average
(per galaxy) stellar consumption rates correlate with central MBH masses
positively for $M\bh\ga10^7\msun$ and negatively for $M\bh\la10^7\msun$. The
volumetric stellar tidal disruption rates are $\sim3\times10^{-5}\pyr\pmpc$ for
MBHs in the mass range of $10^5$--$10^8\msun$ at $z=0$; and the volumetric
stellar consumption rates by MBHs with higher masses are
$\sim10^{-6}\pyr\pmpc$, which can be the stellar tidal disruption rate if the
high-mass BHs are extremely spinning Kerr BHs or the rate of being swallowed
if those BHs are Schwarzschild ones.  The volumetric stellar consumption rates
decrease with increasing redshift, and the decrease is relatively mild for
$M\bh\sim10^5$--$10^7\msun$ and stronger for higher $M\bh$.  Most of the
stellar tidal disruption events (TDEs) at $z=0$ occur in galaxies with mass
$M\gal\la10^{11}\msun$, and about 1\%--2\% of the TDEs can occur in high-mass
galaxies with $M\gal\ga10^{11}\msun$.
\end{abstract}


\section{Introduction}
\label{sec:intro}

Stars can be tidally disrupted (or swallowed whole) by the massive black hole
(MBH) at galactic centers when they move to the vicinity of the MBH
\citep{Hills75}. Studies of stellar tidal disruption events (TDEs) produced by
close encounters with MBHs are of interest in many aspects. The rich phenomena
of TDEs provide not only a probe of much interesting physics in the vicinity of
an MBH (such as the relativistic effects, accretion physics, the interior
structure of the torn stars, and the formation of radio jets via some jetted
TDEs; e.g., \citealt{Komossa15, Alexander17}), but also a constraint on MBH
properties and demographics (such as MBH spins, MBH occupation fractions in
galaxies; e.g., \citealt{Kesden12,Fialkov17}). Stellar tidal disruptions are
one possible channel that could contribute significantly to the present BH mass
in
faint galaxies (e.g., $\la 10^9 L_\odot$) or low-mass MBHs (\citealt{MT99},
hereafter MT99; \citealt{Freitag02,Yu03,Brockamp11,Alexander17}). In addition,
TDEs are among the astrophysical sources of gravitational wave radiation (e.g.,
\citealt{Freitag03}), and synergetic observations of the sources in both
electromagnetic emission and gravitational wave radiation would shed new light
on gravitational wave astrophysics.  In this work, we investigate the expected
TDE rates and their statistical distributions in the real universe.

The rates of TDEs have been estimated both observationally and theoretically.
To date, a few dozen TDE candidates have been observed in the X-ray,
$\gamma$-ray, UV, optical, infrared, and radio bands, and through emission lines, with $\sim$
1--3 new TDE candidates discovered every year (e.g., \citealt{Komossa15,
Alexander17, Stone20}). Based on these observational measurements, the TDE rate
is estimated to be $\sim 10^{-5}$--$10^{-4}\pyr\pgal$ (e.g., \citealt{Donley02,
Esquej08, Maksym10, Wang12, van_Velzen14, Khabibullin14, Auchettl18,
van_Velzen18}), with large statistical uncertainty.
With upcoming time-domain sky surveys, the size of the TDE sample is expected to
expand by orders of magnitude \citep{Komossa15, Stone20}, and thus the rates of
TDEs in different types of galaxies can be determined with an improved
statistics.
The TDE rates have also been estimated analytically or numerically in many
studies (e.g., MT99; \citealt{WM04, SM16}); and the estimates are
typically about a few times $10^{-4}\pyr\pgal$, with the highest disruption
rates occurring in faint cuspy galaxies. It is pointed out by \citet{SM16}
that the analytical estimates appear to predict relatively more events,
compared with those measured by observations. 

The TDE rate in a galaxy depends on how frequently a star can move to a distance
sufficiently close to the MBH to be tidally disrupted. Such a star has a low
orbital angular momentum.  After the initial low-angular-momentum stars in a
stellar system are tidally disrupted (or swallowed whole) by the central MBH
within one orbital period, the TDE rate depends on how frequently high-angular-momentum stars can refill low-angular-momentum orbits to have close
interactions with the MBH. Most of the theoretical estimates of the TDE rates
are mainly based on the refilling of stars onto the low-angular-momentum orbits
by two-body relaxation. Some other mechanisms and phenomena have also been
discussed to contribute to the TDEs or enhance the TDE rates, such as resonant
relaxation (which is shown to enhance the overall rates only modestly;
\citealt{Rauch96, HA06}), the existence of massive perturbers at galactic
centers (which is shown to increase the rate only by a factor of a few;
\citealt{Perets07}), the existence of a secondary MBH at galactic centers
(which is proposed to enhance the rate in a rather short dynamical friction
timescale of $\sim 10^5\yr$ \citealt{Ivanov05,Chen11}), recoiled massive binary
BH merger remnants \citep{Stone11}, and non-spherical galaxy shapes (in which
high-angular-momentum stars can precess onto low-angular-momentum orbits)
(MT99; \citealt{Yu02, Vasiliev13, Vasiliev14}).  In this work, we consider the
effects of triaxial galaxy shapes in enhancing the TDE rates.

The effects of non-spherical gravitational potentials on the TDE rates have
been studied in some previous works (e.g., see MT99; \citealt{Vasiliev13}),
where either an axisymmetric galaxy shape configuration is generally assumed or
a triaxial galaxy shape is assumed to evolve into an axisymmetric shape within
a certain time period (e.g., one hundred times the orbital period) due to the existence of
a central MBH in the stellar system \citep{Merritt98}. In this work, we assume
that the triaxial shape of a galaxy is static, and we do not assume an
evolution of triaxial shapes into axisymmetric shapes, which is plausible, at
least, for the following reasons. (a) A small deviation from a purely
axisymmetric configuration toward a triaxial one may lead to a much larger
reservoir of stars that can move to the vicinity of the central MBHs (cf.\
Fig.~7 in \citealt{CYL20}; hereafter CYL20). (b) The observations of the
nuclear star cluster at our own Galactic center indicate that its shape can be
triaxial even within the radius of the spherical influence of the central MBH
\citep{Feldmeier-Krauseetal17}.

Note that the underlying physics in determining the TDE rates of stars by the
central MBH in a triaxial galaxy has a similarity to that in determining the
dynamical evolution timescale of a massive (hard) binary black holes (BBH)
embedded in a triaxial galaxy merger remnant \citep{Yu02}. The former involves
close interactions of stars with a single MBH, whereas the latter involves
close interactions of stars with a hard BBH system. Both depend on how frequently
low-angular-momentum stars can move to the vicinity of the MBHs.  In CYL20, a
statistical study has been done in how frequently stars can move to the BBH
vicinity in realistic triaxial galaxy distributions, and further to obtain the
distributions of the BBH evolution timescales, and the cosmic massive BBH
populations, and their gravitational wave radiation.  Due to the analogy, in
this work it is straightforward to adopt many similar approaches and galaxy
samples developed in CYL20 to obtain how frequently stars can move to the
vicinity of an MBH to be consumed with considering realistic triaxial galaxy
distributions.  For conciseness, in this paper we refer many relevant aspects
(e.g., galaxy triaxiality shape distributions, galaxy surface brightness
distributions, galaxy stellar mass functions, MBH demographics, the regions of
the phase space in which stars that can precess onto low-angular-momentum
orbits) directly to the relevant sections in CYL20. 

This paper is organized as follows. In Section~\ref{sec:lcdynm}, we briefly
describe the dynamical processes considered in this study that contribute to
TDEs, and review the models for the rate of stellar consumption by the MBH in a
galaxy, with including both the contributions by two-body relaxation and
orbital precession in non-spherical galactic gravitational potentials. In
Section~\ref{sec:framework}, we present the framework for estimating the
statistical distributions of the stellar consumption rates in galaxies with
realistic distributions, as well as the galaxy samples, their distributions,
and the BH demography to be used in the framework.  The main results are
presented in Section~\ref{sec:result}. Conclusions are given in
Section~\ref{sec:concl}.

\section{Loss-cone refilling}
\label{sec:lcdynm}

Consider a stellar system consisting of a central MBH with mass $M\bh$ and some
surrounding stars. A star with mass $m_\ast$ and radius $r_\ast$ that can
travel within a distance of $r<r\rmt \simeq (f_t M\bh/m_\ast)^{1/3} r_\ast$ to
the MBH will be disrupted by the tidal force from the central MBH if
$r\rmt>r\rmswl$, or swallowed whole by the MBH if $r\rmt<r\rmswl$, where
$f_t$ is a factor related to the stellar structure and set to unity in this study,
$r\rmswl\equiv 4GM\bh/c^2$ (with including the general relativistic effect
around a Schwarzschild BH; \citealt{C92}), $G$ is the gravitational constant,
and $c$ is the speed of light. The value of $r\rmswl$ can be smaller/larger
for a prograde/retrograde orbit if the MBH is a spinning Kerr BH.  The
disruption events are accompanied by luminous electromagnetic flares when the
baryonic matter is stripped off the intruding star and accreted onto the MBH
\citep{Hills75,Rees88}.  Following MT99, we define the stellar ``consumption
rate'' as the rate of stars moving to $r<\max(r\rmt,r\rmswl)$ and define the
stellar ``flaring rate'' as the rate of stars moving to $r<r\rmt$ with
$r\rmt>r\rmswl$.
 
Given $(\calE,J)$, where $\calE$ is the specific binding energy (hereafter
energy) and $J$ is the specific angular momentum (hereafter angular momentum)
of a star, the loss cone is defined as the phase space satisfying
\begin{eqnarray}
J\leq J\lc
& \equiv & \max{(\sqrt{2r\rmt^2[\psi(r\rmt)-\calE]},J\rmswl)} \\ \nonumber
& \simeq &
\max{(\sqrt{2GM\bh r\rmt},J\rmswl}),
\label{eq:JlcJswl}
\end{eqnarray}
where $\psi(r)=-\Phi(r)$, $\Phi(r)$ is
the gravitational potential at distance $r$ from the center, $J\rmswl\equiv
4GM\bh/c$ for a Schwarzschild BH (with including the general relativistic
effect; \citealt{C92}).
The loss cone in the context of BBH evolution can be
set by replacing $r\rmt$ by $f_aa$, where $a$ is the separation of the binary
system and $f_a$ is a dimensionless factor $\sim 1$ \citep{Yu02}.

The stars initially in the loss cone will be removed from the loss cone and
consumed by the MBH within one orbital period.  The long-term rate of stellar
consumption by a central MBH is determined by the loss-cone refilling
rate, i.e., how frequently the loss cone is refilled by stars initially outside
the loss cone. In this work, we consider the refilling caused by both two-body
relaxation and orbital precession in non-spherical gravitational potentials.

In non-spherical gravitational potentials, the phase space of $(\calE,J)$ in
which the stars that can precess into the loss cone is described by ``loss
region''. The loss region can be approximated by a ``loss wedge'' [$J_z<J\lc$
with $J<J\rms(\calE)$, where $J_z$ is the component of the $J$ along the
axisymmetric axis] for an axisymmetric system, and approximated by
$J<J\rms(\calE)$ for a triaxial system.  The $J\rms(\calE)$ is the
characteristic angular momentum at energy $\calE$ below which the majority of
stellar orbits are centrophilic. Both the loss wedge in an axisymmetric system
and the loss region in a general triaxial system can be obtained by simulating
stellar motion numerically (CYL20; \citealt{CY14}) or by constructing a
symplectic map in surfaces of section (\citealt{TT97}; MT99).  The rates of
refilling into the loss cone caused by orbital precession constitute the
long-term draining rates of the loss region.
We adopt the relevant analytical results presented in MT99 on the stellar
refilling rates caused by two-body relaxation and the draining rates of a full
loss region, which are summarized in Sections~\ref{sec:lcdynm:lc} and
\ref{sec:lcdynm:drain}, respectively. 

Note that not only is the loss cone in a spherical stellar system refilled
by two-body relaxation, but also the loss wedge/loss region in a non-spherical
stellar system can be refilled by two-body relaxation.
MT99 presents the analysis on the stellar diffusion into the loss wedge by
two-body relaxation in an axisymmetric system.
In this work, by generalizing the analytical result of stellar diffusion
into the loss cone in a spherical stellar system, we include the effect of
stellar diffusion into the loss region by two-body relaxation in a triaxial
system,
and the corresponding analysis is presented in Section~\ref{sec:lcdynm:lc}.

For simplicity, in this study we assume that the MBH at a galactic center is a
single MBH. We ignore the effect of galaxy mergers, starbursts, mass growth,
and the Brownian motion of the central MBH during the long-term period of
stellar consumptions.

\subsection{Two-body relaxation}
\label{sec:lcdynm:lc}

The long-term diffusion rate of stars into the loss cone by two-body relaxation
can be obtained by the steady-state solution of the Fokker-Planck equation. For
simplicity, we assume that all the stars in the stellar system have the same
mass $m_\ast$ and radius $r_\ast$. As for a spectrum of stellar masses and
radii, the derived diffusion rate differs by a factor $\la 2$ (e.g., Appendix
A1 of MT99 and Section 2.3 of \citealt{SM16}).

We define $R\equiv J^2/J\rmc^2(\calE)$ to be the dimensionless squared angular
momentum, where $J\rmc(\calE)$ is the angular momentum of a star at energy
$\calE$ and on a circular orbit and we have $R\le 1$. In a spherical system, the steady-state
diffusion rate of stars into the loss cone obtained by solving the
Fokker-Planck equation is given by (MT99; \citealt{Lightman77,CK78}):
\begin{equation}
F\Lc(\calE)d\calE =
\frac{F^{\max}(\calE)d\calE}{\ln R_0^{-1}(\calE)} =
\frac{4\pi^2 P(\calE)J\rmc^2(\calE)\bar{\mu}(\calE)\bar{f}(\calE)d\calE}
{\ln R_0^{-1}(\calE)},
\label{eq:2brate}
\end{equation}
where $\bar{\mu}(\calE)$ is the orbit-averaged diffusion coefficient,
$\bar{f}(\calE)$ is the `isotropized' distribution function (see Eq.~21 of
MT99), $F^{\max}(\calE)\equiv
4\pi^2\bar{f}(\calE)J\rmc^2(\calE) \bar{\mu}(\calE)P(\calE)$ represents an
estimate of the maximum possible flux through a surface of constant $\calE$ in
phase space.
The $R_0$ in Equation (\ref{eq:2brate}) is
given by
\begin{equation}
R_0(\calE)=R\lc(\calE)\times\left\lbrace
\begin{split}
&\exp(-q)  &q>1 \\
&\exp(-0.186q-0.824\sqrt{q})  &q<1
\end{split},
\right.
\label{eq:r0}
\end{equation}
where $R\lc(\calE)=J^2\lc(\calE)/J\rmc^2(\calE)$ and $q(\calE)\equiv
P(\calE)\bar{\mu}(\calE)/R\lc(\calE)$
(MT99).
The orbit-averaged diffusion coefficient, $\bar{\mu}(\calE)$, is defined by
\begin{equation}
\bar{\mu}(\calE) 
\equiv 
\frac{2}{P(\calE)}\int_{r_-}^{r_+}\frac{\mu dr}{v\rmr}= 
\frac{-2}{P(\calE)}\int_0^{\infty}dv\rmr\frac{dr}{d\psi}\mu(\calE, r),
\label{eq:mubar}
\end{equation}
where $v\rmr$ is the radial velocity satisfying
$v\rmr=\sqrt{2(\psi(r)-\calE)}$, $P(\calE)$ is the radial period of an orbit
with energy $\calE$ and zero angular momentum, and $r_+$ and $r_-$ are the
apocenter and pericenter distances of loss-cone orbits.  The diffusion
coefficient $\mu$ is defined as
\begin{equation}
\mu(\calE,r) \equiv
\left.\langle(\Delta R)^2\rangle/2R\right\vert_{R\rightarrow 0} =
\frac{2r^2\langle(\Delta v\rmt)^2\rangle}{J\rmc^2(\calE)},
\label{eq:mu}
\end{equation}
where $\langle(\Delta v\rmt)^2\rangle$ is the diffusion coefficient for
tangential velocity \citep{BT08}, i.e., 
\begin{equation}
\langle(\Delta v\rmt)^2\rangle = 
\frac{32\pi^2G^2{m_\ast^2}\ln\Lambda}{3}
\left(3I_{\frac{1}{2}}-I_{\frac{3}{2}}+2I_0\right),
\label{eq:dfef}
\end{equation}
with 
\begin{equation}
\begin{split}
&I_0 \equiv \int_0^\calE\bar{f}(\calE')d\calE'\\
&I_{\frac{n}{2}} \equiv [2(\psi(r)-\calE)]^
{-\frac{n}{2}}\int_\calE^{\psi(r)}
[2(\psi(r)-\calE')]^{\frac{n}{2}}\bar{f}(\calE')d\calE'
\end{split}.
\label{eq:iterm}
\end{equation}

In an axisymmetric system, the diffusion rate of stars into the loss wedge by
two-body relaxation, denoted by $F\lw(\calE)$, follows Equations (49) and (50)
in MT99, where $F\lw(\calE)\le F\Lc\full(\calE)$ and
$F\Lc\full(\calE)\equiv 4\pi^2f(\calE)J\lc^2(\calE)$ is the flux of stars per unit
energy consumed by an MBH when the loss cone is full.

In a triaxial system, the diffusion rate of stars into the loss region by
two-body relaxation, denoted by $F\Lr(\calE)$, can be obtained by generalizing
the analysis done for a spherical system, i.e., replacing $J\lc(\calE)$ by
$J\rms(\calE)$ in Equations (\ref{eq:2brate}) and (\ref{eq:r0}).  In a triaxial
system, the diffusion rate of stars into the loss region by two-body relaxation
can contribute to the refilling rate of stars into the loss cone by
$F\Lc\tri(\calE)\equiv \min(F\Lr(\calE),F\Lc\full(\calE))$.  The difference of
$F\Lc\tri(\calE)$ from $F\Lc(\calE)$ is one effect caused by the draining and
depletion of the loss region in a triaxial system as described in
Section~\ref{sec:lcdynm:drain}.

\subsection{Draining a full loss region in non-spherical gravitational potentials}
\label{sec:lcdynm:drain}

In an axisymmetric or a triaxial system, initially the total stellar mass in
the loss region is 
\begin{equation}
M_{\rm lr} \simeq 
m_\ast\int 4\pi^2f(\calE)J\lr^2(\calE)P(\calE)d\calE,
\end{equation}
where $J\lr^2\equiv J\lc J\rms(\calE)$ for an axisymmetric system and
$J\lr^2\equiv J\rms^2(\calE)$ for a triaxial system.
At a sufficiently long time $T$, the draining rate of the loss region due to
orbital precession can be approximated by
\begin{equation}
F\Drain(\calE)d\calE = 
4\pi^2f(\calE)J\lc^2(\calE)
\exp\left[-\frac{T}{P(\calE)}\frac{J\lc^2(\calE)}{J\lr^2(\calE)}
\right]d\calE
\label{eq:Fdrain}
\end{equation}
(cf.\ Eqs.~36 and 52 in MT99).
In this work, the values of $F\Drain(\calE)$ used are set at 
a time $T$ that is the Hubble time of the corresponding redshift.
According to Equations (\ref{eq:2brate}) and (\ref{eq:Fdrain}), the ratio of
the stellar consumption rate caused by orbital precession in the non-spherical
potentials to that caused by two-body relaxation is given by
\begin{equation}
\frac{F\Drain(\calE)}{F\Lc(\calE)} = 
\frac{\ln R_0^{-1}(\calE)}{q(\calE)}
\exp\left[-\frac{T}{P(\calE)}\frac{J\lc^2(\calE)}{J\lr^2(\calE)},
\right]d\calE.
\end{equation}
where ${\ln R_0^{-1}(\calE)}/{q(\calE)}\sim 1$ for $q(\calE)\gg 1$ and ${\ln
R_0^{-1}(\calE)}/{q(\calE)}\sim -\ln R\lc/q(\calE)$ for $q(\calE)\rightarrow 0$.

As presented above, the role of triaxiality of a galaxy is displayed through a
characteristic angular momentum $J\rms(\calE)$, which determines the size of
the loss region. In this work, given the mass density distribution and the
triaxial shape of a host stellar system, we use the Monte Carlo method to
obtain $J\rms(\calE)$ by numerically simulating the motion of stars under the
combined gravitational potential of a central point mass MBH and the galactic
stars (with a triaxial gravitational potential shape) and finding out the
fractions of stars on centrophilic orbits. The detailed numerical methods can
be seen in Section 2.5 in CYL20. Note that a galaxy shape is likely to be close
to being axisymmetric. In practice, given a shape configuration, we use those methods
to estimate both the size of the loss region (by applying some specific
criterion designed for a general triaxial system) and the size of the loss
wedge (by applying some specific criterion designed for an axisymmetric
system). The stars in the loss wedge and in the loss region both contribute to
the stellar reservoir to be consumed by the MBH, and we obtain the draining
rate $F\Drain$ by the maximum of their contributions in Section~\ref{sec:result}.

The stellar consumption rate contributed by the effects of both two-body
relaxation and orbital precession in a non-spherical system can be set by
$F\consp(\calE)\equiv \max(F\Drain(\calE),F\Lc\tri(\calE), F\lw(\calE))$.

\section{Distributions of stellar consumption rates and galaxy/MBH demography}
\label{sec:framework}

\subsection{Distributions of stellar consumption rates}
\label{sec:sample:stat}

We define the energy-integrated stellar consumption rate by $\ndot\equiv \int
F(\calE)d\calE$, which can be expressed as $\Glc$ (with $F\Lc$) to represent
the rate contributed by two-body relaxation in a spherical potential or
$\ndot\consp$ (with $F\consp$) to represent the rate contributed by both
two-body relaxation and orbital precession in non-spherical potentials.
We define the distribution of the stellar consumption rates by
$\Phi_{\ndot}(\ndot,M\bh,z)$ so that $\Phi_{\ndot}(\ndot,M\bh,z)d\ndot dM\bh$
represents the comoving number density of MBHs at redshift $z$ with mass within
the range $M\bh\rightarrow M\bh+dM\bh$ and with stellar consumption rate within
the range $\ndot\rightarrow \ndot+d\ndot$.  We define the galaxy stellar mass
function $n\gal(M\gal,z)$ so that $n\gal(M\gal,z)dM\gal$ represents the comoving
number density of galaxies at redshift $z$ with mass in the range
$M\gal\rightarrow M\gal+dM\gal$.  Similarly, we define the mass function of the
spheroidal components of galaxies (i.e., early-type galaxies and the bulge of
late-type galaxies, which are both called ``bulges'' for simplicity in this
work) $n\bgal(M\bgal, z)$ so that $n\bgal(M\bgal, z)dM\bgal$ represents the
comoving number density of galaxies at redshift $z$ with bulge mass in the range
$M\bgal\rightarrow M\bgal+dM\bgal$. The distribution of the stellar consumption
rate can be given as follows:
\begin{eqnarray}
\Phi_{\ndot}(\ndot,M\bh,z) = & 
\int dM\bgal n\bgal(M\bgal,z)
p\bh(M\bh|M\bgal,z) \nonumber \\
& \times p\gm(\ndot|M\bgal,M\bh,z)
\label{eq:ratedf}
\end{eqnarray}
and
\begin{equation}
n\bgal(M\bgal,z)=
\int dM\gal n\gal(M\gal,z) p\bgal(M\bgal|M\gal,z),
\label{eq:bulgemf}
\end{equation}
where $p\gm(\ndot|M\bgal,M\bh,z)$ is a probability function defined so that
$p\gm(\ndot|M\bgal,M\bh,z)d\ndot$ represents the probability that a bulge with
mass $M\bgal$ at redshift $z$ containing an MBH with mass $M\bh$ has stellar
consumption rate within the range $\ndot\rightarrow \ndot+d\ndot$,
$p\bh(M\bh|M\bgal,z)$ is a probability function defined so that
$p\bh(M\bh|M\bgal,z)dM\bh$ represents the probability that a bulge with mass
$M\bgal$ at redshift $z$ contains an MBH with mass in the range $M\bh\rightarrow
M\bh+dM\bh$, and $p\bgal(M\bgal|M\gal,z)$ is also a probability function defined so
that $p\bgal(M\bgal|M\gal,z)dM\bgal$ represents the probability that a galaxy
with mass $M\gal$ at redshift $z$ contains a bulge with mass in the range
$M\bgal\rightarrow M\bgal+dM\bgal$. We have
$\int p\gm(\ndot|M\bgal,M\bh,z)d\ndot=1$, $\int
p\bh(M\bh|M\bgal,z)dM\bh=1$, and $\int p\bgal(M\bgal|M\gal,z)dM\bgal=1$.
The distributions of $p\bh$ and $p\bgal$ depend on BH demography and galaxy
demography, and the distribution of $p\gm$ depends on the distributions
of the intrinsic structure of galaxies.

The stellar consumption rate $\ndot$ is a function of the parameters describing
the mass density distribution and the shape distribution of host galaxies.  The
mass density distribution and the shape distribution of host galaxies are
likely to evolve with redshifts. In this work, we assume that they do not
evolve with redshift for simplicity. Thus, $p\gm$ is independent of redshift
$z$, i.e., $p\gm(\ndot|M\bgal,M\bh,z)= p\gm(\ndot|M\bgal,M\bh)$. When the
redshift evolution information of the mass density distribution and the shape
distribution is available with future observations, $p\gm$ can be updated
accordingly.  

In a galaxy sample, suppose that there are $N$ galaxies with bulge mass and
MBH mass $(M\bgal,M\bh)$ that have the stellar consumption rate
$\ndot_{i|M\bgal,M\bh}$, $(i=1,2,...,N)$ (with $T$ being set to the Hubble time
at the corresponding redshift in Eq.~\ref{eq:Fdrain} for the draining in the
non-spherical case). The probability function $p\gm$ can be determined through
\begin{eqnarray}
& & p\gm(\ndot|M\bgal,M\bh) \nonumber \\
& = &
\frac{1}{N}\sum_{i=1}^{N}\delta(\ndot-\ndot_{i|M\bgal,M\bh}) \nonumber \\
& = & \frac{1}{N}\frac{d}{d\ndot}\sum_{i=1}^{N}H(\ndot-\ndot_{i|M\bgal,M\bh}),
\label{eq:pgamma}
\end{eqnarray}
where $H(x)$ is a step function defined by $H(x)=1$ if $x>0$ and $H(x)=0$ if
$x\le 0$.
Applying Equation~(\ref{eq:pgamma}) into Equation~(\ref{eq:ratedf}), we have
\begin{eqnarray}
\Phi_{\ndot}(\ndot,M\bh,z) & = &
\frac{d}{d\ndot} \int dM\bgal  [ n\bgal(M\bgal,z) p\bh(M\bh|M\bgal,z) \nonumber \\
 & & \times \frac{1}{N}\sum_{i=1}^{N}H(\ndot-\ndot_{i|M\bgal,M\bh}) ] .
\label{eq:ratedf1}
\end{eqnarray}

We define an enhancement factor $f\enhance$ to characterize the enhancement in
the average stellar consumption rate due to both two-body relaxation and
orbital precession in non-spherical potentials as compared to the
rate due to two-body relaxation in spherical potentials as follows:
\begin{equation}
f\enhance(M\bh,z)\equiv \frac{\langle \Gfuel\rangle(M\bh,z)}{\langle \Glc\rangle(M\bh,z)},
\label{eq:fenhance}
\end{equation}
where the expectation of the consumption rate $\langle \ndot\rangle$ for MBHs
with mass $M\bh$ at redshift $z$ is defined by
\begin{equation}
\langle \ndot\rangle(M\bh,z) \equiv
\frac{\int \ndot\Phi_{\ndot}(\ndot,M\bh,z) d\ndot}{\int\Phi_{\ndot}(\ndot,M\bh,z)d\ndot}.
\label{eq:xpec}
\end{equation}

We define the differential volumetric 
consumption rate over MBH mass $M\bh$ at redshift $z$ by
\begin{equation}
\frac{d\GVinte(M\bh,z)}{d\log M\bh}\equiv 
(M\bh\ln 10)\int\ndot\Phi_{\ndot}(\ndot,M\bh,z) d\ndot,
\label{eq:rvol_diff}
\end{equation}
where $\GVinte (M\bh,z)$ is the integrated volumetric
consumption rate
over a range of MBH mass lower than $M\bh$ at redshift $z$.

Note that it is straightforward to generalize the above framework to define the
stellar consumption rate as a function of galaxy mass $M\gal$, instead of
$M\bh$. In Section~\ref{sec:result}, we will present the volumetric stellar
consumption rates not only as a function of $M\bh$, but also as a function of
$M\gal$ (with omitting a similar formalism as above for simplicity).

\subsection{Galaxy/MBH demography}
\label{sec:sample}

We summarize the galaxy samples, the distributions of galaxy properties, and
the MBH demography to be used in this work below, which share many
similarities to those developed in CYL20, as mentioned in
Section~\ref{sec:intro}.

To model the statistical distributions of stellar consumption rates, we use the
observational samples of early-type galaxies in the ATLAS$^{\rm 3D}$ survey
\citep{Cappellari11} and those in \citet{Lauer07} to generate a mock galaxy
sample with a large size of $\sim 10^7$ galaxies, as described in Sections 4.1
and 4.2 in CYL20). Each galaxy in the mock sample is characterized by its
generated surface brightness profile and mass-to-light ratio.  The mock galaxy
sample with the large size covers a sufficiently large variety in galaxy
properties. Note that one important parameter to characterize the inner stellar
distribution of a galaxy is the inner slope $\gamma$ in the surface brightness
profile of a galaxy ($I(R)\propto R^{-\gamma}$ as $R\rightarrow 0$; see Eq.~38
in CYL20), and one of our results shown in Section~\ref{sec:result} is to
investigate the dependence on $\gamma$. 

Note that by constructing the mock galaxy sample based on the observational
galaxy samples above, we have implicitly assumed that the distribution of the
combined observational galaxy sample is representative of the distribution of
the galaxies in the real universe in the parameter space of the five Nuker-law
parameters and the mass-to-light ratio.

Regarding the intrinsic triaxial shape distributions of galaxies, we use the
observational results summarized from the ATLAS$^{\rm 3D}$ survey and the Sloan
Digital Sky Survey. The triaxial shape of a galaxy in the mock sample is
generated randomly from the observational galaxy triaxiality distributions
presented in \citet{Padilla08} and \citet{Weijmans14} (See more details in
Fig.~2 and Section 4.1 in CYL20). Compared to the shape configurations shown in
\citet{Padilla08}, those shown in \citet{Weijmans14} are relatively closer to
axisymmetric. In this work, the results are obtained with both of the shape
configurations for comparison.

The galaxy stellar mass function $n\gal(M\gal,z)$ used in this work is obtained
from \citet{Behroozi19} (see Figure 3 and Section 4.3.1 in CYL20). The
probability function of $p\bh$ in Equation (\ref{eq:ratedf}) can be set through
the BH--bulge relations obtained from different works in the literature (see
more details in Eq.~46, Figure 5, Table 1, and Section 4.4 in CYL20).  
The probability function of $p_*$ in Equation
(\ref{eq:bulgemf}) is set through galaxy demography on the fractions of
different types of galaxies \citep{Behroozi19} and their bulge-to-total stellar
mass ratios (see more details in Section 4.4 in CYL20).

\subsection{The stellar consumption rate library and its application}
\label{sec:sample:sample:libr}

To evaluate the distributions of the stellar consumption rate statistically (see
Eq.~\ref{eq:ratedf}), we construct a consumption rate library based on the mock
galaxy sample. In the library, we set the MBH mass so that $\log(M\bh/\msun)$ is
in the range from 5 to 10 with an interval of 0.2. For each MBH mass considered,
we select 500,000 galaxies randomly from the mock galaxy sample. Then based on
the description in Section~\ref{sec:lcdynm}, the stellar distribution in each
galaxy is modeled, and the stellar consumption rates $\Glc$ and 
$\Gfuel$ are estimated.

The stellar consumption rate library is then utilized to obtain the
distributions of the stellar consumption rates in the realistic universe (see
Eq.~\ref{eq:ratedf1}) by a Monte-Carlo method.  We evaluate the distribution
function of $\Phi_{\ndot}(\ndot,M\bh,z)$ at the mesh points of $\log(M\bh/\msun)$ (in
the range from 5 to 10 with an interval of 0.2), same as the mesh points in the
stellar consumption rate library.  For each given MBH mass, we set the
integration range of $\log (M\gal/\msun)$ to be within $[7,13]$ in Equation
(\ref{eq:bulgemf}), and divide the integration range into 60 bins with an
interval $\Delta\log M\gal=0.1$. Within each integration bin, we generate 2000
galaxies based on the stellar mass function $n\gal(M\gal,z)$.  For each of the
2000 galaxies, we generate its bulge mass $M\bgal$, based on the probability
distribution of $p_*$. For each galaxy (e.g., denoted by $i$) characterized by
$(M\bgal,M\bh)$ in a bin, we search the stellar consumption rate library to
find out the system with the same MBH mass $M\bh$ and with the bulge mass being
closest to $M\bgal$, and then assign its stellar consumption rate in the
library to galaxy $i$ generated in the bin, denoted by $\ndot_i$. In this way,
the distributions of the stellar consumption rates are evaluated by the sum
over all the galaxy mass bins, by using Equations
(\ref{eq:ratedf})--(\ref{eq:ratedf1}).

\section{Results}
\label{sec:result}

In this section, we show the simulation results on how the rates of stellar
consumption by central MBHs are affected by the inclusion of the
loss-region draining in non-spherical potentials, and how they depend on
MBH masses and inner stellar distributions of the host galaxies. We also present
the differential and the integrated volumetric consumption rates at different
MBH masses, different redshifts, and different galaxy masses. Some detailed
results are illustrated by
Figures~\ref{fig:mock_hist}--\ref{fig:mock_rvol_Mgal_m9}, where the MBHs are
assumed to be Schwarzschild BHs, and $m_*$ and $r_*$ are set to be the solar mass
and radius. The results are obtained by using the MBH mass versus the galaxy
bulge mass relation of \citet{Kormendy13}; i.e., ``KH13b'' in Table 1 of CYL20.
In Figures~\ref{fig:mock_rvol}--\ref{fig:mock_rvol_Mgal_m9}, we also present the
median of the results obtained with all the different BH--host galaxy relations
listed in Table 1 of CYL20 and the standard deviations around the median.

\subsection{Stellar consumption rates per galaxy}

\begin{figure*}[!htb]
\centering
\includegraphics[width=0.85\textwidth]{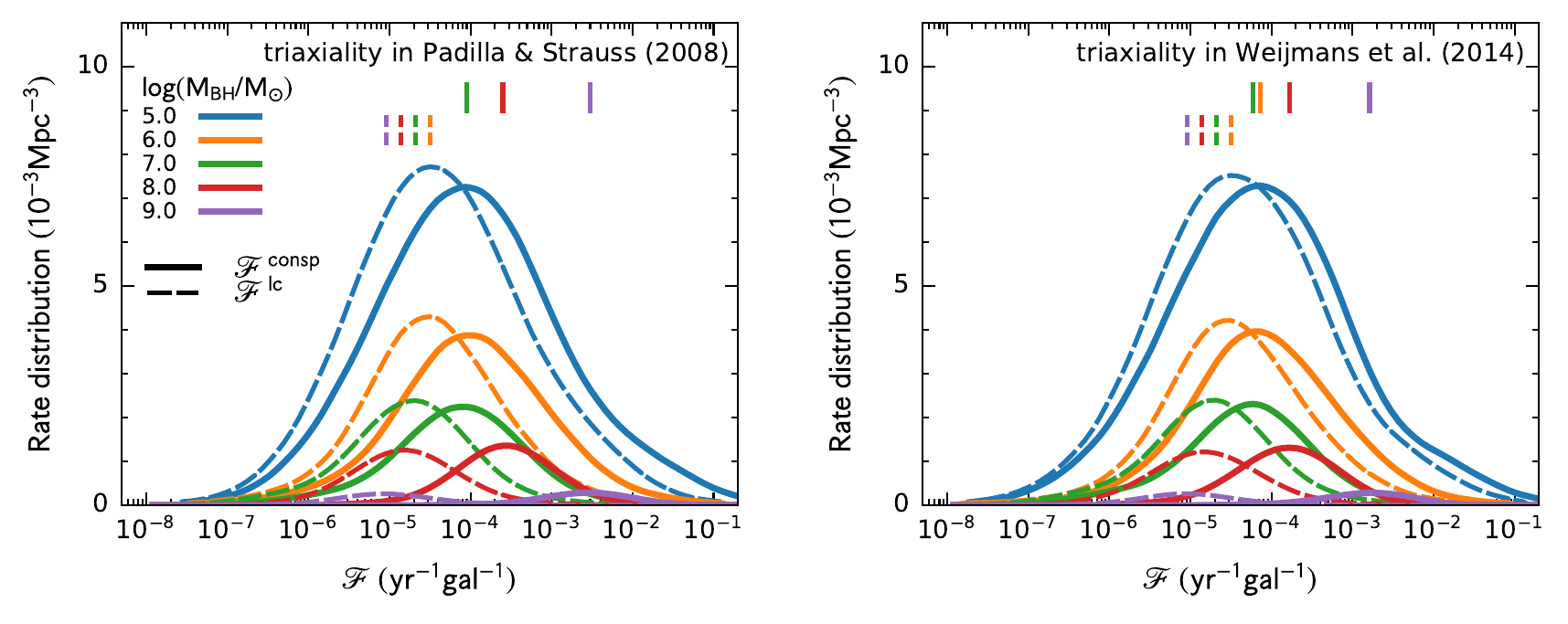}
\caption{ 
Rate distributions of stellar consumption by MBHs (see Eq.~\ref{eq:ratedf}),
$\ndot M\bh(\ln 10)^2\Phi_{\ndot}(\ndot, M\bh, z=0)$, and their dependence on
MBH masses and galaxy shape distributions, where $\ndot$ is $\Glc$ (dashed line;
due to two-body relaxation) or $\Gfuel$ (solid line; due to the effects of
non-spherical galaxy shapes). The different colors represent different MBH
masses,
and the peak position of each distribution is marked by a vertical short line
with its corresponding color and line style. In the left panel, the galaxy shape
distribution of \citet{Padilla08} is adopted; and in the right panel, the galaxy
shape distribution of \citet{Weijmans14} is adopted.  As seen from the figure,
given an MBH mass, the $\Gfuel$ value at the peak location is larger than the
corresponding $\Glc$ value at the peak location. The $\Glc$ peak value shifts
leftward with increasing MBH masses $M\bh$, while the $\Gfuel$
peak value shifts rightward with increasing MBH masses $M\bh$ from $10^7\msun$ to higher masses.
The offsets
between the distributions of $\Glc$ and the $\Gfuel$ increase with increasing
MBH masses.
For $M\bh\sim 10^5$--$10^6\msun$, the offset between the $\Glc$
and $\Gfuel$ peak values is $\sim$ a factor of 3; and it becomes more than one order of
magnitude for $M\bh\simeq 10^8\msun$. Compared to the offsets shown in the
left panel, those in the right panel are only slightly smaller, which indicates
that the loss-region draining (obtained with galaxy shape configurations
close to being axisymmetric in \citet{Weijmans14}) is only slightly less
effective.  The large offsets for high-mass MBHs ($M\bh \sim 10^8$--$10^9\msun$)
suggest that triaxial galaxy shapes are effective in increasing the rates of
stellar tidal disruption by the high-mass MBHs if the MBHs are extremely spinning
Kerr BHs or increasing the rates of stars swallowed by the MBHs if the MBHs are
Schwarzschild BHs. See Section~\ref{sec:result}.} \label{fig:mock_hist}
\end{figure*}

\begin{figure*}[!htb]
\centering
\includegraphics[width=0.85\textwidth]{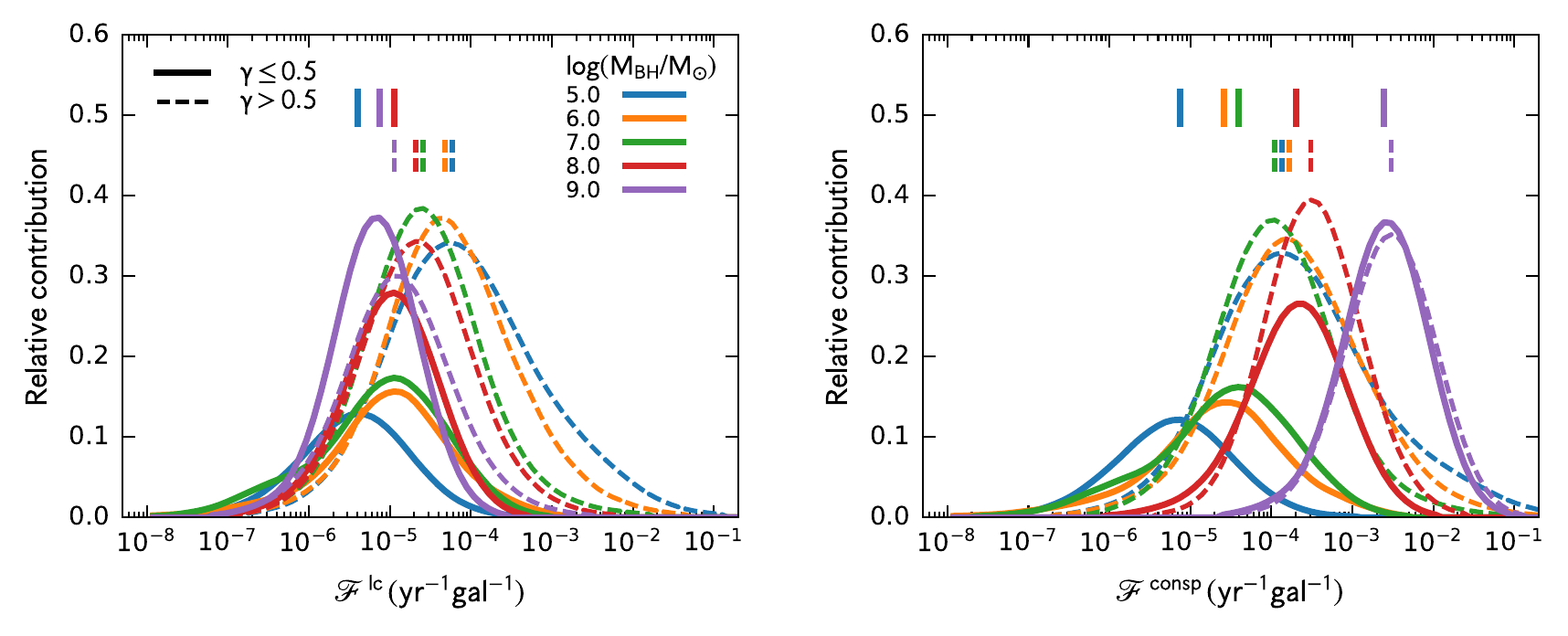}
\caption{Dependence of the stellar consumption rate distributions on the inner
slope of the surface brightness profiles of galaxies $\gamma$. The vertical axis
shows the relative contributions of ``core'' galaxies (with $\gamma\le 0.5$;
solid curves) and ``cusp'' galaxies ($\gamma>0.5$; dashed curves) to the
distributions of the stellar consumption rate of all the galaxies, defined by
$\frac{(\ln 10)\ndot\Phi_{\ndot}(\ndot,M\bh,z=0)|_{\gamma\le 0.5}} {\int
\Phi_{\ndot}(\ndot,M\bh,z=0)d\ndot}$ (solid curves) and $\frac{(\ln
10)\ndot\Phi_{\ndot}(\ndot,M\bh,z=0)|_{\gamma>0.5}} {\int
\Phi_{\ndot}(\ndot,M\bh,z=0)d\ndot}$ (dashed curves), respectively, where
$\ndot$ corresponds to $\Glc$ in the left panel and 
$\Gfuel$ in the right panel. The peak position of each curve
is marked by a short vertical line with its corresponding color and line style.
The galaxy shape distribution of \citet{Padilla08} is adopted in this figure. As
seen from the shapes and the peak locations of the curves in the left panel, the
average rates of stellar consumptions due to two-body relaxation are higher in
galaxies with high $\gamma$ than those in galaxies with low $\gamma$, which is
significant especially for low-mass MBHs (e.g., $M\bh\simeq 10^5\msun$) and
becomes less significant as the MBH mass increases. Similarly, the shapes and
the peak locations of the distributions in the right panel (for $\Gfuel$ due to
the effects of non-spherical galaxy shapes) show that
the average stellar consumption rates are also significantly higher in galaxies
with high $\gamma$ for $M\bh\sim 10^5$--$10^7\msun$. See Section~\ref{sec:result}. } \label{fig:mock_hist_gamma}
\end{figure*}

Figure~\ref{fig:mock_hist} plots the distributions of the (per-galaxy) stellar
consumption rates by central MBHs (see Eq.~\ref{eq:ratedf}), and their
dependence on different MBH masses, different mechanisms of refilling stars into
the loss cone, and different adopted galaxy shape distributions. We only show
the distributions for the MBH population at redshift $z=0$. The distributions at
other redshifts have different magnitudes, proportional to the number densities
of the MBHs at the corresponding redshifts.  In each panel, the distributions
shown by the dashed curves represent the cases in which the loss-cone refilling
is due to two-body relaxation, while the distributions shown by the solid curves
represent the cases in which the loss-cone refilling is due to the
combined effects of two-body relaxation and orbital precession in triaxial
galaxy potentials. The offsets between the two sets of
distributions reveal the enhancement in the estimated stellar consumption rates
caused by considering the draining of the loss region in
non-spherical galaxies. As seen from
the figure, the peaks of the distributions of $\Glc$ (dashed lines) shift
toward lower $\Glc$ values with increasing $M\bh$. In contrast,
the peaks of the distributions of $\Gtot$ (solid curves) shift toward higher
values of $\Gtot$ (from $\sim 10^{-4}$ to $\sim
3\times 10^{-3}\pyr\pgal$) with increasing $M\bh$ from $10^7\msun$ to $10^9\msun$. 
In Table~\ref{tab:fig1}, we list the values of $\log[\ndot/(\pyr\pgal)]$ at the
peaks of the distributions, as well as the logarithms of the mean values of
$\ndot$. The average stellar consumption rates of $\Glc$ (due to two-body
relaxation) have a negative correlation with MBH masses $M\bh$ (as low-mass MBHs
tend to be located in galaxies with steeper inner slopes; see also
Fig.~\ref{fig:mock_hist_gamma}). The average stellar consumption
rates of $\Gtot$ (including the effects of triaxial galaxy shapes) range from
$\sim 5\times 10^{-4}$ to $\sim 6\times 10^{-3}\pyr\pgal$, and have a
negative correlation with $M\bh$ for $M\bh\la 10^7\msun$ but a positive
correlation for higher masses due to the effects of triaxial galaxy shapes.
For smaller MBHs $M\bh\sim
10^5$--$10^7\msun$, the offset between the peak locations of the solid and the
dashed curves is $\sim$0.5~dex along the horizontal axis. For MBH mass $M\bh\simeq 10^8\msun$, the offset in the peak
locations can be more than one order of magnitude if the triaxial galaxy shape
distribution of \citet{Padilla08} is adopted (left panel).
The offset is
slightly smaller but still considerable if the galaxy shape distribution of
\citet{Weijmans14}, which prefers axisymmetric configurations, is adopted (right
panel).  For MBHs with even higher masses, the offset becomes larger. The rates
contribute little to the observable flare rates of tidal disruptions if the
high-mass MBHs are Schwarzschild BHs, as most of the stars may be swallowed
whole by the MBHs. However, if the MBHs are extremely spinning Kerr BHs, stars
can still be tidally disrupted due to general relativistic effects (e.g.,
\citealt{Kesden12,MB20}), and the figure suggests that the triaxial galaxy
shapes are effective in increasing the stellar tidal disruption rates,
especially for high-mass MBHs ($M\bh\sim 10^8$--$10^9\msun$).

\setcounter{table}{4}
\begin{deluxetable*}{ccccccccccccccccccccccccccccc}
\setlength{\tabcolsep}{0.05in}
\tablenum{1}
\tablewidth{0pt}
\tablecaption{The peak and the mean (per-galaxy) stellar consumption rates.}
\tablehead{
  \colhead{$\log(M\bh/\msun)$}
&&\colhead{$\log\Glc|\rmpeak$}
&&\colhead{$\log\langle \Glc \rangle$}
&&\colhead{$\log\Gfuel|\rmpeak$ }
&&\colhead{$\log\langle \Gfuel\rangle$}
&&\colhead{$\log\Gfuel|\rmpeak$ }
&&\colhead{$\log\langle \Gfuel\rangle$}
\\
\cline{7-9}
\cline{11-13}
    \colhead{}
&& \multicolumn{3}{c}{}
&& \multicolumn{3}{c}{\citet{Padilla08}}
&& \multicolumn{3}{c}{\citet{Weijmans14}}
}
\startdata
5.0 & & -4.50 & & -3.03 & & -4.05 & & -2.55 & & -4.14 & & -2.66 \\
6.0 & & -4.50 & & -3.51 & & -4.05 & & -2.94 & & -4.14 & & -3.10 \\
7.0 & & -4.68 & & -3.97 & & -4.05 & & -3.33 & & -4.23 & & -3.59 \\
8.0 & & -4.86 & & -4.21 & & -3.60 & & -3.15 & & -3.78 & & -3.41 \\
9.0 & & -5.04 & & -4.54 & & -2.52 & & -2.25 & & -2.79 & & -2.53 \\
\hline
\enddata
\tablecomments{The peak and the mean values of the stellar consumption rates for
the distributions shown in Figure~\ref{fig:mock_hist}. The stellar consumption
rate $\ndot$ is in units of $\pyr\pgal$. The results obtained by applying the
galaxy triaxiality distributions from \citet{Padilla08} and \citet{Weijmans14}
are both listed in the table.}
\label{tab:fig1}
\end{deluxetable*}

We investigate the role played by the inner stellar distribution of a galaxy in
regulating the stellar consumption rates and show the results in
Figure~\ref{fig:mock_hist_gamma}. As mentioned in Section~\ref{sec:sample}, one
important parameter to characterize the inner stellar distribution of a galaxy
is the inner slope $\gamma$ in the surface brightness profile of a galaxy.  We
divide the host galaxies into ``core'' galaxies with $\gamma\le 0.5$ and
``cusp'' galaxies with $\gamma> 0.5$, and show their relative contributions to
the total stellar consumption rate distributions by the solid and the dashed
curves, respectively.  The relative contributions shown in
Figure~\ref{fig:mock_hist_gamma} are defined by the ratios of $\frac{(\ln
10)\ndot\Phi_{\ndot}(\ndot,M\bh,z=0)|_{\gamma\le 0.5}} {\int
\Phi_{\ndot}(\ndot,M\bh,z=0)d\ndot}$ (solid curves) and $\frac{(\ln
10)\ndot\Phi_{\ndot}(\ndot,M\bh,z=0)|_{\gamma>0.5}} {\int
\Phi_{\ndot}(\ndot,M\bh,z=0)d\ndot}$ (dashed curves), where $\ndot$ is $\Glc$ in
the left panel and is $\Gfuel$ in the
right panel.  The results shown in Figure~\ref{fig:mock_hist_gamma} are obtained
by adopting the galaxy shape distribution of \citet{Padilla08}, and adoption of
the shape distribution of \citet{Weijmans14} reveals a similar trend. As seen
from the peak locations and the shapes of the curves in the left panel, the
average rates of stellar consumptions due to two-body relaxation are higher in
the galaxies with steeper inner slopes, especially for low-mass MBHs.
In the
right panel, the average stellar consumption rates are also higher in galaxies
with high $\gamma$ for $M\bh\sim
10^5$--$10^7\msun$. For MBHs with higher masses, the difference in the
peak locations of the distributions of core galaxies and cusp galaxies becomes
negligible.

\begin{figure*}[!htb]
\centering
\includegraphics[width=0.85\textwidth]{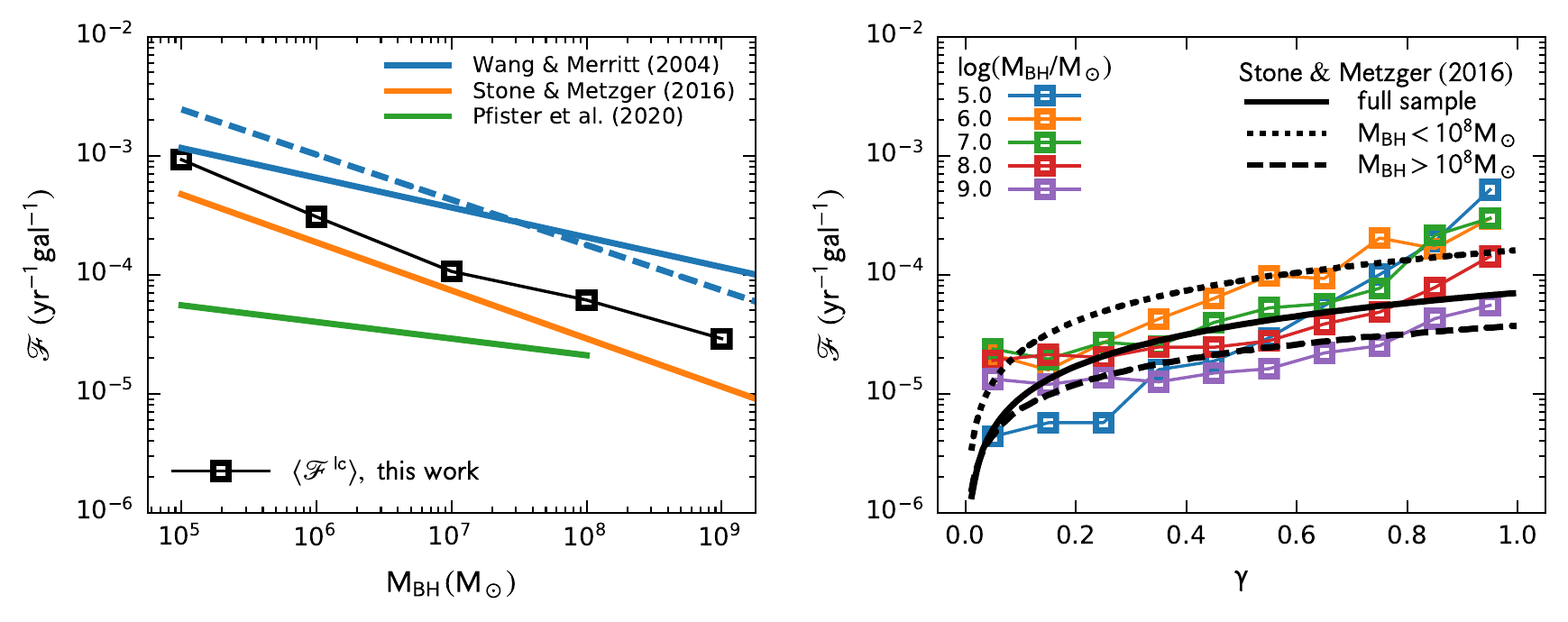}
\includegraphics[width=0.85\textwidth]{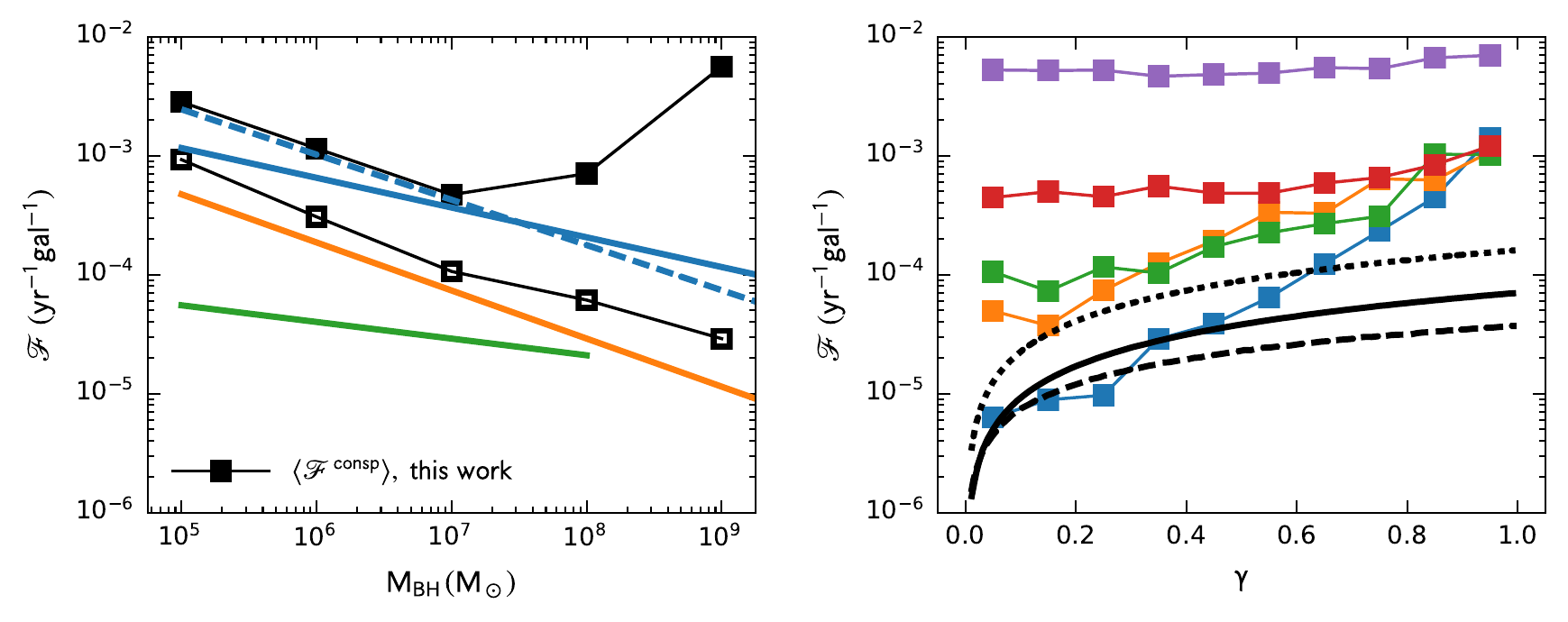}
\caption{
Dependence of the stellar consumption rate by MBHs $\ndot$ on the MBH mass
$M\bh$ (left panels) and the inner slope of the surface brightness profile
$\gamma$ (right panels). In the upper left panel, the open squares represent
the mean stellar consumption rates due to two-body relaxation $\Glcmean$
obtained in our study; and the other colored lines represent the results obtained
in some other works, which correspond to the singular isothermal sphere
model-based relation of \citet{WM04} obtained with the $M\bh$--$\sigma$
relation of \citet{MF01} (blue solid line) and with the $M\bh$--$\sigma$ relation of
\citet{MM13} (blue dashed line), the fitting result of \citet{SM16} in which the MBH
masses follow the $M\bh$--$\sigma$ relation of \citet{MM13} (orange solid line), and
the fitting result of \citet{Pfister20} to their mock galaxy sample (green
solid line), respectively. In the upper right panel, the open squares represent
$\Glcmean$ obtained in our study, and each square represents the average rate
for the host galaxies with $\gamma$ being in some specific ranges, i.e.,
[0,0.1], [0.1,0.2], ..., and [0.9,1.0]; and the different colors represent
different MBH masses. The panel also shows the fitting results of \citet{SM16}
to their full galaxy sample (black solid line), to their subsample with MBH masses
below the mass criterion for tidal disruption of (solar-type) stars
(black dotted line), and to their subsample with MBH masses above the mass criterion (black dashed line), respectively.  In
the lower panels, we show the mean stellar consumption rate due to the effects
of triaxial galaxy shapes, $\Gfuelmean$, by filled squares.
As seen from the figure, $\Glcmean$
correlates negatively with $M\bh$, consistent with most of the other works
(e.g., \citealt{WM04, SM16, Pfister20}); and it correlates positively with
$\gamma$, and the correlation becomes stronger for low-mass MBHs.
The $\Gfuelmean$ has a negative correlation with $M\bh$ for
$M\bh\la 10^7\msun$, and the correlation becomes positive for
$M\bh\ga 10^7\msun$. 
The correlation between $\Gfuelmean$ and $\gamma$ is positive
for $10^5\la M\bh\la 10^7\msun$ and becomes negligible
for larger MBHs.
See Section~\ref{sec:result}.
} \label{fig:mock_corr} \end{figure*}

Figure~\ref{fig:mock_corr} shows both the mean stellar consumption rate due to
two-body relaxation $\Glcmean$ (open squares) and the mean consumption rate
due to triaxial galaxy shapes $\Gfuelmean$ (filled squares; lower panels) as a
function of $M\bh$ (left panels) and $\gamma$ (right panels).  To get the
relation of $\Glcmean$ or $\Gfuelmean$ with $\gamma$, we calculate the
distributions of stellar consumption rates in galaxies with different $\gamma$
ranges from 0 to 1 with an interval of 0.1, similarly as done for the
distributions of core/cusp galaxies in Figure~\ref{fig:mock_hist_gamma} (but
with the above different $\gamma$ range cuts).
In the left panels, we also show the results of some previous works for
comparison, i.e., the result obtained based on the singular isothermal sphere
model of \citet{WM04} by applying the $M\bh$--$\sigma$ relation of \citet{MF01}
(blue solid line) or the $M\bh$--$\sigma$ relation of \citet{MM13} (blue dashed line), the
fitting result of \citet{SM16} in which the MBHs follow the $M\bh$--$\sigma$
relation of \citet{MM13} (orange solid line), and the fitting result of
\citet{Pfister20} to their mock galaxy sample (green solid line). In the right
panels, we also show the fitting result of \citet{SM16} to their full galaxy
sample (black solid line), to their subsample with MBH masses below the mass
criterion for tidal disruption of (solar-type) stars (black dotted line), and to that with MBH
masses above the mass criterion (black dashed line), respectively. 
As seen from the upper panels, the amplitude of our mean stellar consumption
rate due to two-body relaxation $\Glcmean$ in spherical potentials is well
consistent with that of \citet{SM16}, as their galaxy
sample from \citet{Lauer07} is one part of our observational sample used to
generate the mock galaxy sample. In
our work, $\Glcmean$ has a negative correlation with $M\bh$, consistent with
most of the previous works (e.g., see \citealt{WM04, SM16, Kochanek16,
Pfister20}). As can be seen from the lower panels,
the inclusion of the effects of triaxial galaxy shapes not only improves the
average stellar consumption rate but also changes its correlation with both
$M\bh$ and $\gamma$.

In the upper left panel of Figure~\ref{fig:mock_corr}, the slopes in the
$\log(\Glc)$--$\log(M\bh)$ correlation vary among the different works, e.g.,
-0.25 and -0.38 for \citet{WM04} where the $M\bh$--$\sigma$ relations of
\citet{MF01} and \citet{MM13} are adopted, respectively, -0.404 for the fitting
to the full galaxy sample in \citet{SM16} and -0.247/-0.223 for the fitting to
their core/cusp galaxy subsamples, and -0.14 for the galaxy samples with MBH
masses below the mass criterion for tidal disruption of (solar-type) stars and
considering
the effects of nuclear star clusters in some of the galaxies in
\citet{Pfister20}.  A linear fit to the five data points obtained in our work
gives a slope of -0.37.  In the lower left panel, our result obtained with
the inclusion of the effects of triaxial galaxy shapes shows that $\Gfuelmean$
has a negative correlation with $M\bh$ for $M\bh\la 10^7\msun$, and the
correlation becomes positive for $M\bh\ga 10^7\msun$.  The $\Gfuelmean$ is
larger than $\Glcmean$ by a factor of $\sim$3 for $10^5\la M\bh\la 10^7\msun$, and
by more than one and two orders of magnitude for $M\bh\simeq 10^8\msun$ and
$10^9\msun$, respectively.

In the upper right panel of Figure~\ref{fig:mock_corr}, our result of
$\Glcmean$ shows a positive correlation with $\gamma$, qualitatively consistent
with \citet{SM16}. The correlation is stronger for low MBH masses: at high
$\gamma$ (e.g.,$\sim 1$), $\Glcmean$ decreases with increasing $M\bh$; and at
low $\gamma$ (e.g.,$\sim 0$), $\Glcmean$ first increases and then decreases
with increasing $M\bh$ (see also \citealt{WM04} for the transition between
these two trends as $\gamma$ decreases, i.e., Figure~3 and Section~4
therein).
The lower right panel shows that there exists a positive correlation between
$\Gfuelmean$ and $\gamma$ for $10^5\la M\bh\la 10^7\msun$; however, the
correlation becomes quite weak for larger MBHs.

The dichotomy behavior of the $\Gfuelmean$--$M\bh$ and the
$\Gfuelmean$--$\gamma$ correlations for low-mass MBHs ($M\bh\la 10^7\msun$) and
those high-mass ones ($M\bh\ga 10^7\msun$) is because in the two MBH mass
ranges the dominant flux of stellar consumption have different origins of the
stellar low-angular-momentum orbits, though the stars all move into the
loss cone from the loss region.
In spherical galaxies, the stellar consumption flux is dominated by stellar
diffusion into the loss cone by two-body relaxation. In triaxial galaxies,
stars initially inside the loss region are drained at a timescale of
$P(\calE)J\lr^2(\calE)/J\lc^2(\calE)$. Given a characteristic energy of
stellar orbits (e.g., with apocenter distances being at the influential radius
of an MBH), for low-mass MBHs, the stars initially inside the loss region are
depleted in a relatively shorter timescale, and then the stellar consumption is
dominated by the flux of the stars initially outside the loss region 
diffusing into the loss region due to two-body relaxation (and then precessing
into the loss cone).  For high-mass MBHs, stars initially inside the loss
region are depleted in a relatively longer timescale, e.g., longer than a
Hubble time, and the stellar consumption is dominated by the loss-region
draining rate of the stars initially inside the loss region.  The above
different origins can serve as an explanation for the dichotomy trends from a
statistical perspective, though this might not be the case in some individual
systems.

\begin{figure*}[!htb]
\centering
\includegraphics[width=0.45\textwidth]{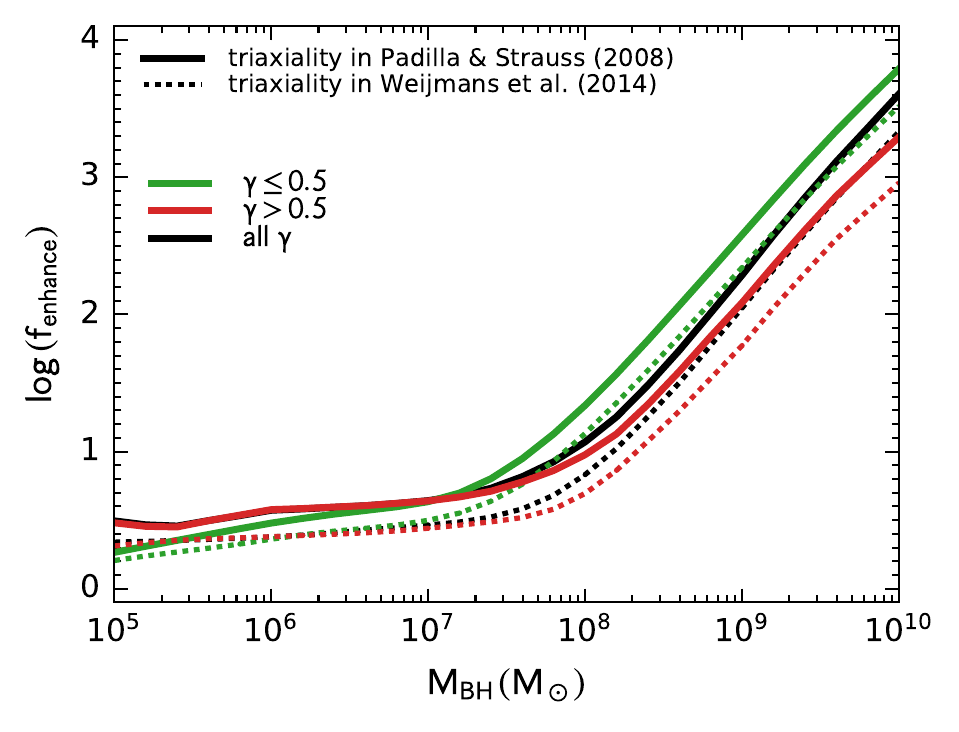}
\caption{
The enhancement factor $f\enhance(M\bh,z=0)$ (Eq.~\ref{eq:fenhance}) as a
function of central MBH mass $M\bh$, describing the enhancement of the stellar
consumption rate due to adding the effects of triaxial galaxy shapes. The
solid and the
dotted curves show the results obtained by using the galaxy shape distribution
of \citet{Padilla08} and that of \citet{Weijmans14}, respectively.  The green
and the red curves correspond to the results obtained for ``core'' galaxies
($\gamma\leq 0.5$) and ``cusp'' galaxies ($\gamma> 0.5$), respectively, and the
black curves represent the results obtained for all the ``core'' and ``cusp''
galaxies.  As seen from the figure, the enhancement factor $f\enhance$
increases with increasing MBH mass $M\bh$. For $M\bh\la 10^7\msun$, the
increasing trend is relatively weak; when $M\bh\ga 10^8\msun$, the trend becomes
much stronger. For $M\bh\sim 10^5$--$10^7\msun$, the enhancement
is a factor of $\sim$3 if the galaxy shape distribution of
\citet{Padilla08} is adopted; for $M\bh\simeq 10^8\msun$, the enhancement can
be about 
one order of magnitude. For $M\bh$ higher than a few times of $10^7\msun$, the
enhancement is larger in core galaxies than that in cusp galaxies. The
enhancement obtained by using the galaxy shape distribution of \citet{Padilla08}
is slightly larger than that obtained by using the distribution of
\citet{Weijmans14}. See Section~\ref{sec:result}.} \label{fig:mock_xpec}
\end{figure*}

Figure~\ref{fig:mock_xpec} plots the enhancement factor $f\enhance(M\bh,z=0)$
(Eq.~\ref{eq:fenhance}) as a function of MBH mass $M\bh$, to show the
enhancement in stellar consumption rates due to the inclusion of effects of
triaxial galaxy shapes. We also show $f\enhance$ for core galaxies and cusp
galaxies by the green and the red curves, respectively, to see how the rate
enhancement is affected by the inner stellar distribution of host galaxies.  As
seen from the figure, the enhancement is a factor of $\sim 3$ for MBHs with
mass $M\bh\la 10^7\msun$, which is dominated by stars refilling the
loss region due to two-body relaxation (and then precessing into the loss
cone).  The enhancement factor $f\enhance$ increases with the increase of MBH
mass $M\bh$, and can be up to about one order of magnitude at $M\bh\simeq
10^8\msun$ and more than two orders of magnitude at $M\bh\simeq 10^9\msun$.
Compared to the results obtained by using the galaxy shape distribution of
\citet{Padilla08}, the enhancement factor obtained by adopting the galaxy shape
distribution of \citet{Weijmans14} (which prefers axisymmetric configurations)
is slightly smaller, e.g., by $\sim 0.3$~dex at $M\bh\simeq 10^8\msun$.  For
MBHs with mass greater than a few times of $10^7\msun$, the enhancement in core
galaxies is larger than that in cusp galaxies, as the consumption rate due to
two-body relaxation in core galaxies is smaller than that in cusp galaxies (see
Figure~\ref{fig:mock_hist_gamma}).

\begin{figure*}[!htb]
\centering
\includegraphics[width=0.45\textwidth]{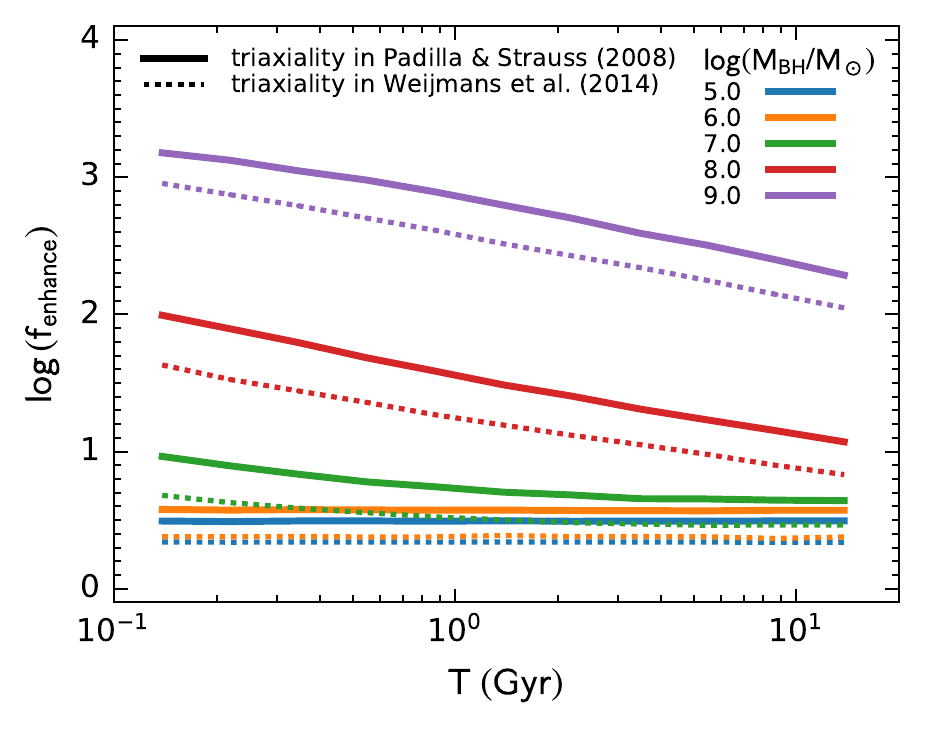}
\caption{
The enhancement factor $f\enhance$ as a function of the draining time $T$
(i.e., the time duration that the loss region has been drained; see $T$ in
Equation~\ref{eq:Fdrain}). The different
colors correspond to different MBH masses. The solid and dotted curves show the
results obtained by using the galaxy shape distribution of \citet{Padilla08} and
that of \citet{Weijmans14}, respectively. As seen from the figure, $f\enhance$
decreases mildly with increasing $T$ for MBHs with mass $M\bh\sim
10^8$--$10^9\msun$ ($\sim 0.5$~dex from $T\sim 1$ to $10\gyr$); and it is
almost flat at $T\sim$1--10$\gyr$
for MBHs with mass $M\bh\sim 10^5$--$10^7\msun$.
The mild change of $f\enhance$ with $T$ suggests that
the statistical
results of the stellar consumption rates obtained in this work should not be
affected significantly by the assumption of the draining time $T$ being set to
the Hubble time of the corresponding redshift. 
See Section~\ref{sec:result}.
}
\label{fig:mock_xpec_Tdrain} \end{figure*}

As mentioned in Section~\ref{sec:lcdynm:drain}, we obtain the draining rate of
a loss region at a time $T$ that is the Hubble time of the corresponding redshift
in Equation (\ref{eq:Fdrain}).  However, the host galaxies may be younger, or
they may have gone through multiple major mergers during their lifetime so that
their distributions have been changed, or the structures with different mass
scales may form at different cosmic time. To see how the stellar consumption
rate in the galaxies with different lifetimes are affected by triaxial galaxy
shapes, we show the enhancement factor $f\enhance$ as a function of the
draining time $T$ for MBHs with different masses in
Figure~\ref{fig:mock_xpec_Tdrain}.  As seen from the figure, from $T=1\gyr$ to
the Hubble time at $z=0$, the enhancement factor changes little for $M\bh\sim
10^5$--$10^7\msun$, and decreases by $\sim 0.5$~dex for $M\bh\sim
10^8$--$10^9\msun$.  If a galaxy formed or experienced its last major merger at
$z\sim 1$, the enhancement factor at $T=8\gyr$ (the cosmic time since $z\sim
1$) is higher than the enhancement factor at the Hubble time at $z=0$ only by
$\sim 0.2$~dex for $M\bh\sim 10^8$--$10^9\msun$, and is almost the same as the
enhancement factor at the $z=0$ Hubble time for $M\bh\sim 10^5$--$10^7\msun$.
Thus the statistical results of the stellar consumption rates obtained in this
work (for normal galaxies) should not be affected significantly by the
assumption of $T$ being set to the Hubble time of the corresponding redshift.
Note that for high-mass MBHs, though the loss-region draining rate at a given
$\calE$ shown in Equation (\ref{eq:Fdrain}) decrease exponentially with
increasing $T$, the mild decrease of $f\enhance$ with increasing $T$ shown in
Figure~\ref{fig:mock_xpec_Tdrain} is a consequence of the facts that
$f\enhance$ is contributed by the draining rates at different $\calE$, and that
the draining timescales $P(\calE)J\lr^2(\calE)/J\lc^2(\calE)$ are different at
different $\calE$ (which can be longer than the relevant timescale, e.g., the
Hubble timescale, for low $\calE$). 

\subsection{Volumetric stellar consumption rates}

\begin{figure*}[!htb]
\centering
\includegraphics[width=0.85\textwidth]{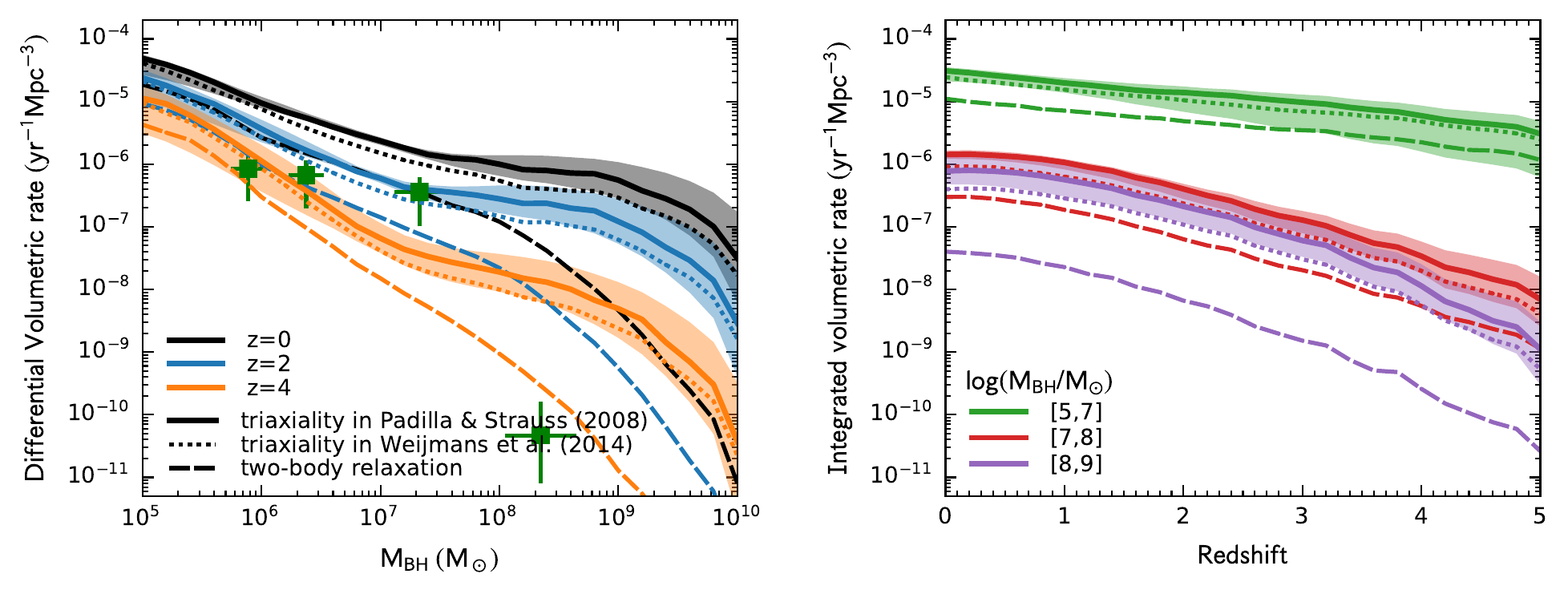}
\includegraphics[width=0.85\textwidth]{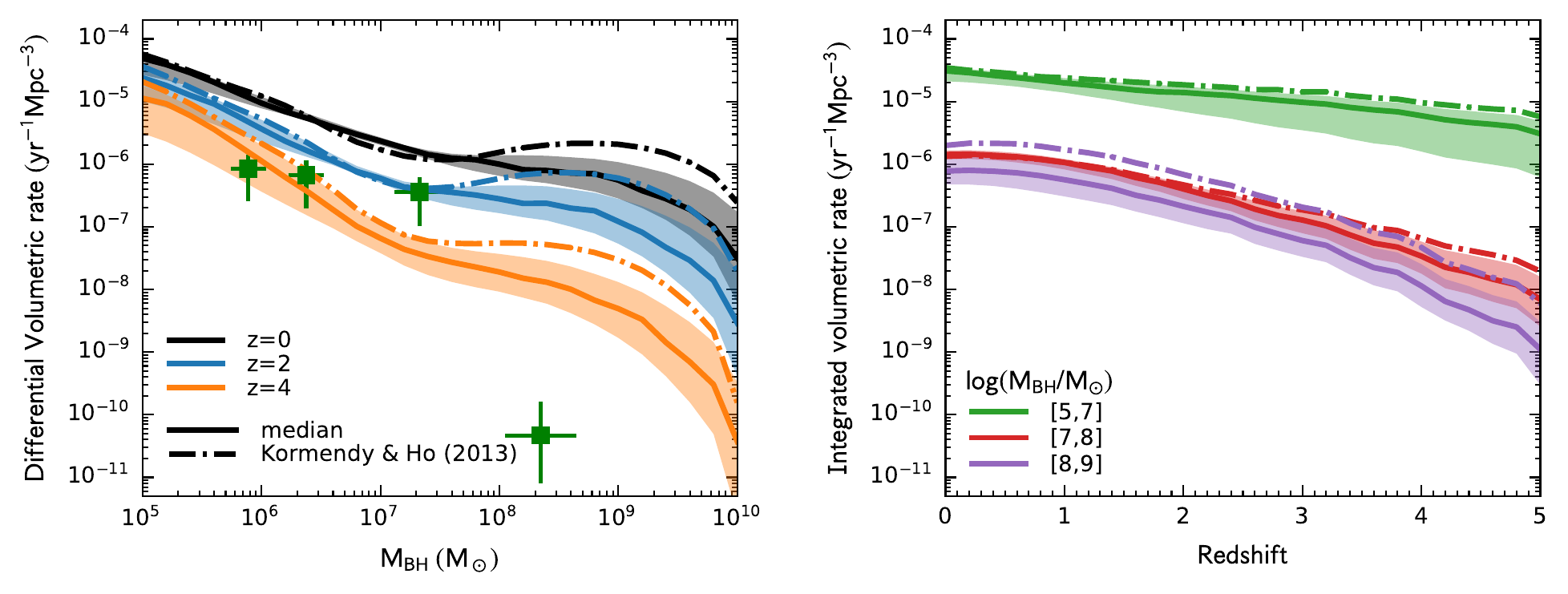}
\caption{
The differential volumetric stellar consumption rate as a function of MBH mass,
$\GVdiff(M\bh,z)$ (see Eq.~\ref{eq:rvol_diff}; left panels) and the integrated
volumetric stellar consumption rate as a function of redshift (right panels).
The integrated volumetric stellar consumption rate are obtained by an
integration of the differential volumetric stellar consumption rate over the
ranges of $\log(M\bh/\msun)$=[5,7], [7,8], and [8,9], respectively.  In the
upper left panel, the black, blue, and orange curves correspond to the results at
redshift $z=0$, 2, and 4, respectively.  The dashed curves represent $\GlcVdiff$
(upper left panel) and their integrated rates (upper right panel) due to the
two-body relaxation alone, and the solid and the dotted curves represent
$\GtotVdiff$ (left panels) and their integrated rates (right panels) obtained by
adding the effects of the loss-region draining in triaxial galaxies, with the
galaxy shape distributions of \citet{Padilla08} and \citet{Weijmans14},
respectively. Each curve (except for the dotted-dashed curve in the bottom
panels) is the median of the corresponding results obtained with the different
BH-host galaxy relations listed in Table~1 of CYL20, and the shaded region
around the solid curve represents the standard deviation of the logarithm of the
results around the corresponding median.  The dotted-dashed curves in the bottom
panels represent $\GtotVdiff$ (bottom left panel) and their integrated rates
(bottom right panel) obtained with the MBH mass versus the galaxy bulge mass
relation of \citet{Kormendy13} and the galaxy shape distributions
\citet{Padilla08}, for a comparison with the medians (solid curves).  
In the left panels, the green squares are the volumetric TDE flaring
rate estimated from a recent compilation of the TDE candidates
shown in Figure 10
of \citep{van_Velzen18}, where the number of the observational sources to
determine each data point is 5, 4, 2 and 1 from left to right, respectively.
As seen from the upper left panel, 
the median $\GfuelVdiff$ is higher than the median $\GlcVdiff$
by a factor of $\sim$3 for MBHs with mass $M\bh\la 10^7\msun$ and by a factor of
$\sim$8 at $M\bh\simeq 10^8\msun$ if the galaxy shape distribution of
\citet{Padilla08} is adopted. The enhancement is slightly smaller if 
the galaxy shape distribution of \citet{Weijmans14} is adopted. The
enhancement is about two orders of magnitude at $M\bh\simeq 10^9\msun$.
As seen from the figure, the stellar consumption rates decrease with increasing
redshifts, which is relatively mild for $M\bh\sim 10^5$--$10^7\msun$ and
stronger for higher $M\bh$. The figure suggests that the triaxial galaxy shapes
are effective in increasing the stellar tidal disruption rates.
See Section~\ref{sec:result}.} \label{fig:mock_rvol} \end{figure*}

Figure~\ref{fig:mock_rvol} plots the differential volumetric consumption rate
$\GlcVdiff$ and $\GfuelVdiff$ (Eq.~\ref{eq:rvol_diff}) as a function of MBH mass
$M\bh$ at the three different redshifts $z=$0, 2, and 4 (left panels) and the
integrated volumetric stellar consumption rate as a function of redshift (right
panels). We obtain the rates by applying the different BH--host galaxy relations
(listed in Table 1 of CYL20), and for each case we show the medians of the rates
in the figure (solid, dotted, dashed curves).  The rates obtained with the MBH
mass versus the galaxy bulge mass relation of \citet{Kormendy13} are shown by
the dotted-dashed curves in the bottom panels, which are generally higher than
the corresponding median results (solid curves) at $M\bh\ga 10^8\msun$. For the
solid curves, we use the shaded region to illustrate the standard deviation of
the logarithms of the rates obtained with the different BH--host galaxy
relations around the medians.  As seen from the figure, the volumetric stellar
consumption rates decrease with increasing redshifts, and the decrease is
relatively mild for $M\bh\sim 10^5$--$10^7\msun$ and stronger for higher $M\bh$.
Note that the redshift evolution of the integrated volumetric consumption rate
$\GVlcinte$ in our study is different from that of \citet{Kochanek16} in the
detailed shape or slope of the evolution curve, though the normalizations at
redshift $z=0$ are consistent. We find that such a difference may be caused by
the MBH mass function and its redshift evolution in the two works,
as the volumetric consumption rate is proportional to the MBH mass
function. An improvement in determining the redshift evolution of
the MBH mass function can help to reveal the redshift evolution of the
stellar consumption rates, and vice versa.
Consideration of the effects of triaxial galaxy shapes can cause the median
volumetric stellar consumption rate to be enhanced by a factor of $\sim$3 for
MBHs with mass $M\bh\la 10^7\msun$ if the galaxy shape distribution of
\citet{Padilla08} is adopted.  This enhancement increases to a factor of
$\sim$8 and to two orders of magnitude at $M\bh\simeq 10^8\msun$ and $\simeq
10^9\msun$, respectively. If the galaxy shape distribution of
\citet{Weijmans14} is adopted, the enhancement is slightly smaller.
We list the integrated volumetric consumption rate at $z=0$ in
Table~\ref{tab:fig4}. For the case of adopting the galaxy shape distribution of
\citet{Padilla08}, the median integrated volumetric consumption rate contributed
by central MBHs within the mass range of $10^5$--$10^7\msun$ due to two-body
relaxation [$\GVlcinte(M\bh=10^7\msun)-\GVlcinte(M\bh=10^5\msun)$] and that
obtained by considering the effect of galaxy triaxial shapes
[$\GVfuelinte(M\bh=10^7\msun)-\GVfuelinte(M\bh=10^5\msun)$] are $1.1\times
10^{-5}\pyr\pmpc$ and $3.1\times
10^{-5}\pyr\pmpc$ at $z=0$, respectively; the corresponding median rates
contributed by MBHs within the mass range of $10^7$--$10^8\msun$ are
$3.0\times 10^{-7}\pyr\pmpc$ and
$1.4\times
10^{-6}\pyr\pmpc$, respectively. The corresponding median rates contributed by
MBHs within the mass range of $10^8$--$10^9\msun$ are $4.0\times
10^{-8}\pyr\pmpc$ and $7.8\times
10^{-7}\pyr\pmpc$, respectively; and the rate obtained with the BH--host galaxy
relation in \citet{Kormendy13} can be up to $2.0\times 10^{-6}\pyr\pmpc$. If the MBHs with $M\bh\sim 10^8$--$10^9\msun$ are
extremely spinning Kerr BHs, the upper mass limit of the MBH that can tidally
disrupt a star can increase from $\sim 10^8\msun$ to $\sim 10^9\msun$ (e.g.,
\citealt{MB20,Kesden12}), the rates of stellar disruption by those BHs can be
effectively improved by the effects of the galaxy triaxiality, which enables a
significant event rate with respect to those for low-mass MBHs especially at low
redshifts (e.g., $\sim$1\%--10\% of the rates for low-mass MBHs).

In the left panels of Figure~\ref{fig:mock_rvol}, we also
plot the volumetric TDE flaring rate estimated from a recently compiled sample
of TDE candidates with black hole mass measurements \citep{van_Velzen18}, where
the number of the observational sources to determine each data point is 5, 4, 2,
and 1 from left to right.  It appears that the observational
constraints are smaller than the model predictions by a factor of up to ten or
more.  One may note that the sample incompleteness, or unknown selection effects
and MBH occupation fractions in low-mass galaxies, or some other effects may
lead to large errors in the observational constraints, which hinders a direct
comparison between observations and model results \citep[see also][]{SM16}.
Accumulation of more observational sources and better understanding of 
various effects involved in the estimates in future will help to put
constraints on the model.

\setcounter{table}{4}
\begin{deluxetable*}{ccccccccccccccccccc}
\setlength{\tabcolsep}{0.05in}
\tablenum{2}
\tablewidth{0pt}
\tablecaption{The integrated volumetric stellar consumption rates at $z=0$.}
\tablehead{
    \colhead{$\log(M\bh/\msun)$}
&& \multicolumn{5}{c}{$\int \frac{d\GVlcinte}{d\log M\bh} d\log M\bh$}
&& \multicolumn{5}{c}{$\int \frac{d\GVfuelinte}{d\log M\bh} d\log M\bh$}
&& \multicolumn{5}{c}{$\int \frac{d\GVfuelinte}{d\log M\bh} d\log M\bh$}
\\
   \colhead{}
&& \multicolumn{5}{c}{}
&& \multicolumn{5}{c}{\citet{Padilla08}}
&& \multicolumn{5}{c}{\citet{Weijmans14}}
\\
\cline{3-7}
\cline{9-13}
\cline{15-19}
  \colhead{}
&&\colhead{KH13}
&&\colhead{median}
&&\colhead{stddev}
&&\colhead{KH13}
&&\colhead{median}
&&\colhead{stddev}
&&\colhead{KH13}
&&\colhead{median}
&&\colhead{stddev}
}
\startdata
$[5,7]$
&& $1.1\times 10^{-5}$ && $1.1\times 10^{-5}$ && $^{+0.03}_{-0.11}$~dex
&& $3.5\times 10^{-5}$ && $3.1\times 10^{-5}$ && $^{+0.06}_{-0.17}$~dex
&& $2.6\times 10^{-5}$ && $2.5\times 10^{-5}$ && $^{+0.04}_{-0.18}$~dex \\
$[7,8]$
&& $2.2\times 10^{-7}$ && $3.0\times 10^{-7}$ && $^{+0.04}_{-0.07}$~dex
&& $1.3\times 10^{-6}$ && $1.4\times 10^{-6}$ && $^{+0.06}_{-0.07}$~dex
&& $8.1\times 10^{-7}$ && $9.1\times 10^{-7}$ && $^{+0.05}_{-0.06}$~dex \\
$[8,9]$
&& $6.7\times 10^{-8}$ && $4.0\times 10^{-8}$ && $^{+0.16}_{-0.15}$~dex
&& $2.0\times 10^{-6}$ && $7.8\times 10^{-7}$ && $^{+0.22}_{-0.22}$~dex
&& $1.1\times 10^{-6}$ && $4.0\times 10^{-7}$ && $^{+0.23}_{-0.21}$~dex \\
\hline
\enddata
\tablecomments{The integrated volumetric stellar consumption rates at $z=0$
shown in Figure~\ref{fig:mock_rvol}, in unit of $\pyr\pmpc$. The MBH
mass range of the integration is listed in the column of ``$\log(M\bh/\msun)$''.
The ``KH13'' column lists the rates obtained with the MBH mass versus
the galaxy bulge mass relation of \citet{Kormendy13}. The ``median''
column lists the median of the rates obtained
with the different BH-host galaxy relations listed in Table~1 of CYL20;
and the ``stddev'' column lists the standard deviation of the
logarithm of the rates around the median in unit of dex.
The results obtained by
applying the galaxy triaxiality distributions from \citet{Padilla08}
and \citet{Weijmans14} are both listed.
}
\label{tab:fig4}
\end{deluxetable*}

Note that the MBH mass in Figures~\ref{fig:mock_hist}--\ref{fig:mock_xpec} is
obtained by adopting the MBH mass versus the galaxy bulge mass relation of
\citet{Kormendy13}, which is generally higher than most of the other BH--host
galaxy relations shown in Figure 5 and Table 1 of CYL20.  Adoption of most of
the other different BH--host galaxy relations can result in an increase of the
per-galaxy stellar consumption rates.  For most of the other different BH--host
galaxy relations, our calculations show that the peak locations shown in
Figure~\ref{fig:mock_hist} will shift toward higher stellar consumption rates
by $\sim$0--0.5~dex, together with wider distribution shapes.
As the galaxy comoving number density decreases significantly with increasing
galaxy masses at the high-mass end, adoption of the other different BH--host
galaxy relations results in the decrease of the volumetric stellar consumption
rates for high-mass BHs ($M\bh\ga 10^8\msun$; e.g., by a factor of a few or up
to one order of magnitude), as illustrated by the comparison of the
dotted-dashed curves and the solid curves in Figure~\ref{fig:mock_rvol}.  
(See Figure 19 of CYL20 for another example of the effect caused by the
scattering of the BH--host galaxy relations, in the predicted stochastic
gravitational wave radiation background of cosmic supermassive binary black
holes.)

In the lower left panel of Figure~\ref{fig:mock_rvol}, there appears a ``bump''
feature in the dotted-dashed curve in the MBH mass range between a few times
$10^7\msun$ and $10^{10}\msun$ (obtained with the MBH mass to bulge mass
relation of \citet{Kormendy13}). The appearance of the bump is because within
this mass range, the per-galaxy stellar consumption rate obtained with
including the effects of triaxial galaxy shapes $\Gfuel$ is an increasing
function of MBH mass $M\bh$, whereas the comoving number density of MBHs is a
decreasing function of $M\bh$. The shape of the curve results from the
combination of these two opposite trends, and the preference for a heavier MBH
at a given galaxy mass set by the MBH mass to bulge mass relation of
\citet{Kormendy13} makes the bump feature prominent.

\begin{figure*}[!htb]
\centering
\includegraphics[width=0.85\textwidth]{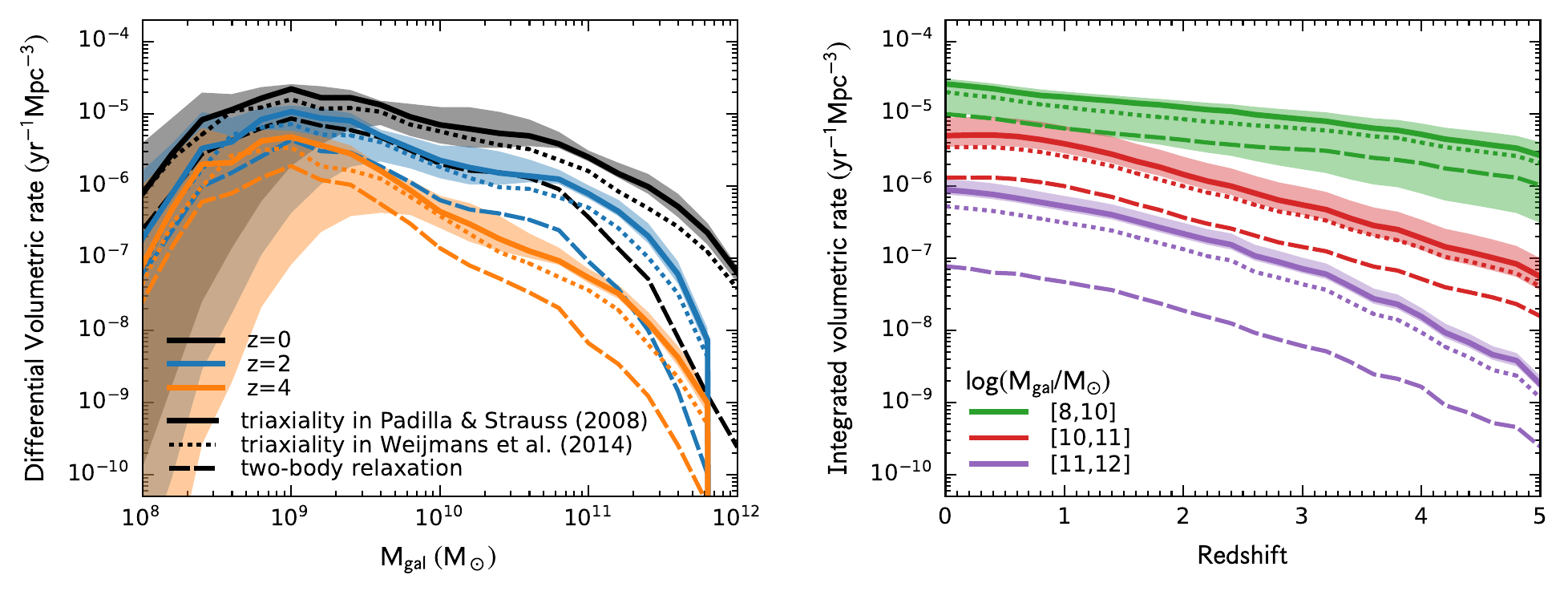}
\includegraphics[width=0.85\textwidth]{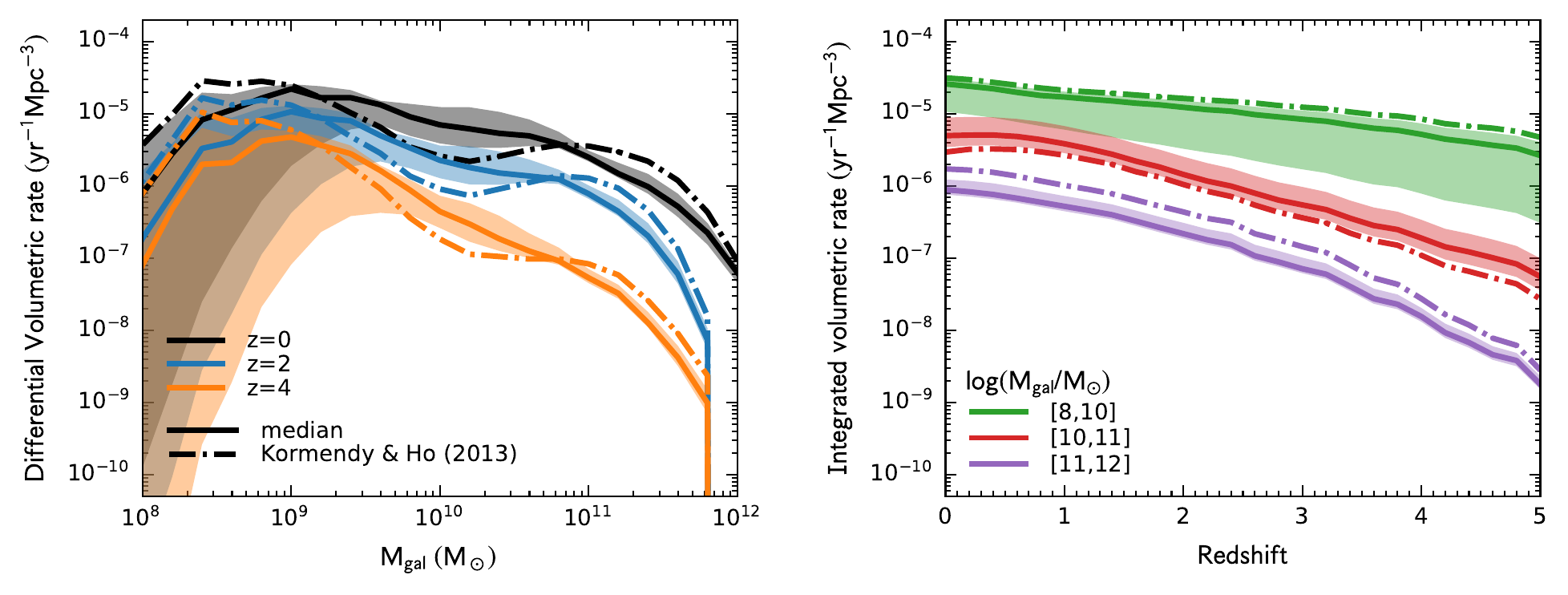}
\caption{The volumetric stellar consumption rate as a function of galaxy
stellar mass $M\gal$. The curves, colors, and texts have meanings similar
to those in Figure~\ref{fig:mock_rvol}, except that the rates in this
figure are expressed as a function of $M\gal$, instead of $M\bh$.
} \label{fig:mock_rvol_Mgal} 
\end{figure*}

\begin{figure*}[!htb]
\centering
\includegraphics[width=0.85\textwidth]{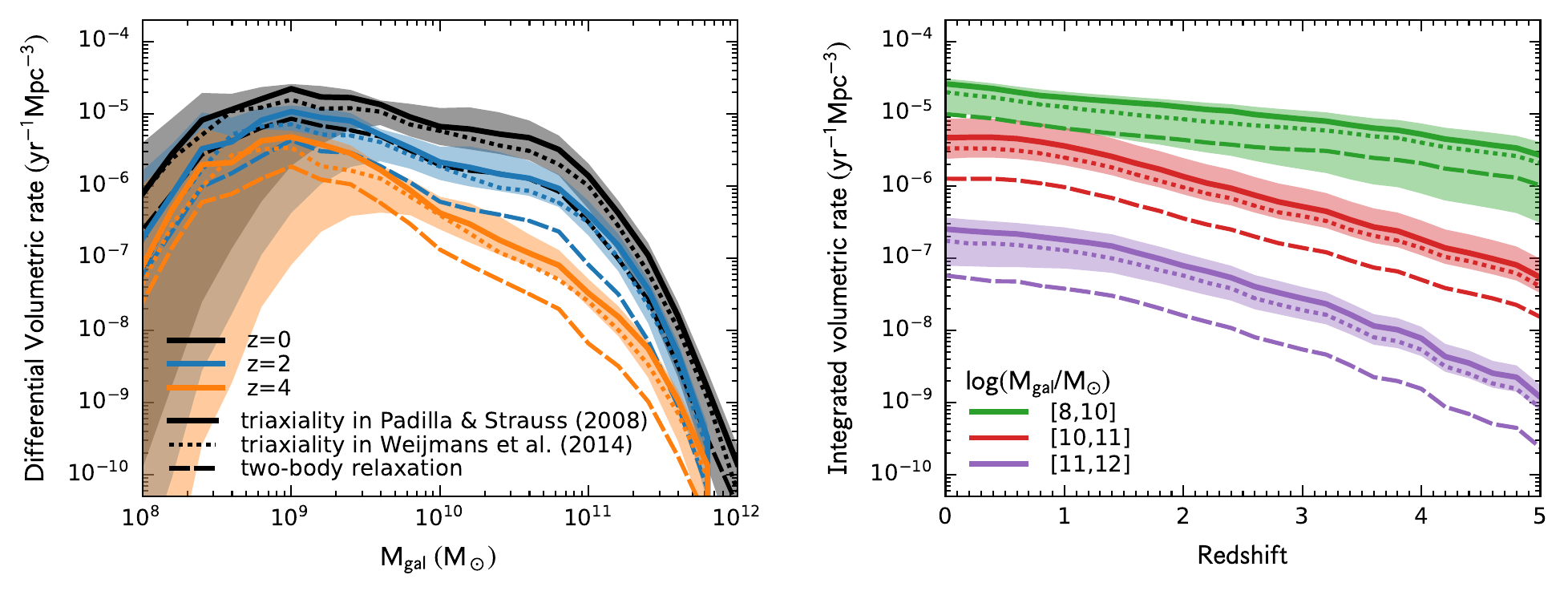}
\includegraphics[width=0.85\textwidth]{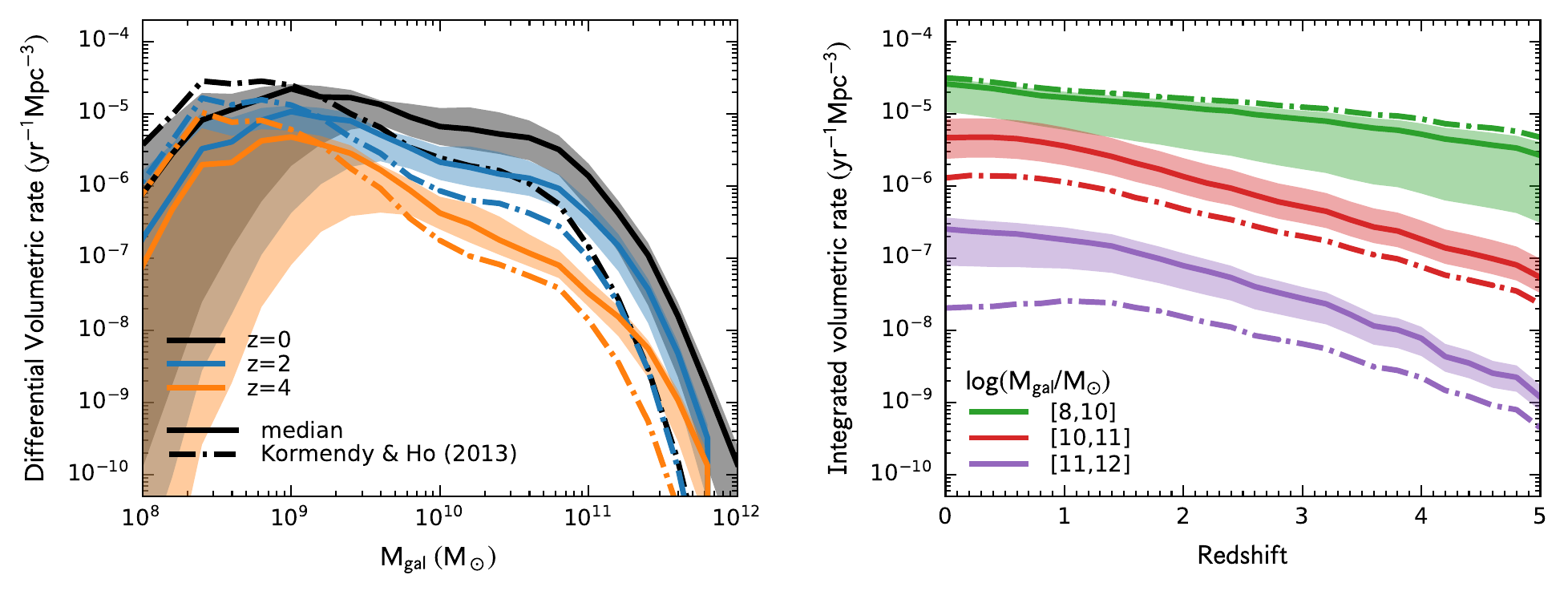}
\caption{Similar to Figure~\ref{fig:mock_rvol_Mgal}, but excluding galaxies with
MBH mass $M\bh>10^8\msun$. The rates in high-mass galaxies ($M\gal\sim
10^{11}$--$10^{12}\msun$) decrease significantly, compared to those in
Figure~\ref{fig:mock_rvol_Mgal}. The volumetric stellar disruption rates in
bright galaxies ($\sim 3\times 10^{-7}\pyr\pmpc$ at $z=0$ for $M\gal\ga 10^{11}\msun$) are about
1\% of the stellar tidal disruption rates in all the galaxies at $z=0$. }
\label{fig:mock_rvol_Mgal_m8} \end{figure*}

\begin{figure*}[!htb]
\centering
\includegraphics[width=0.85\textwidth]{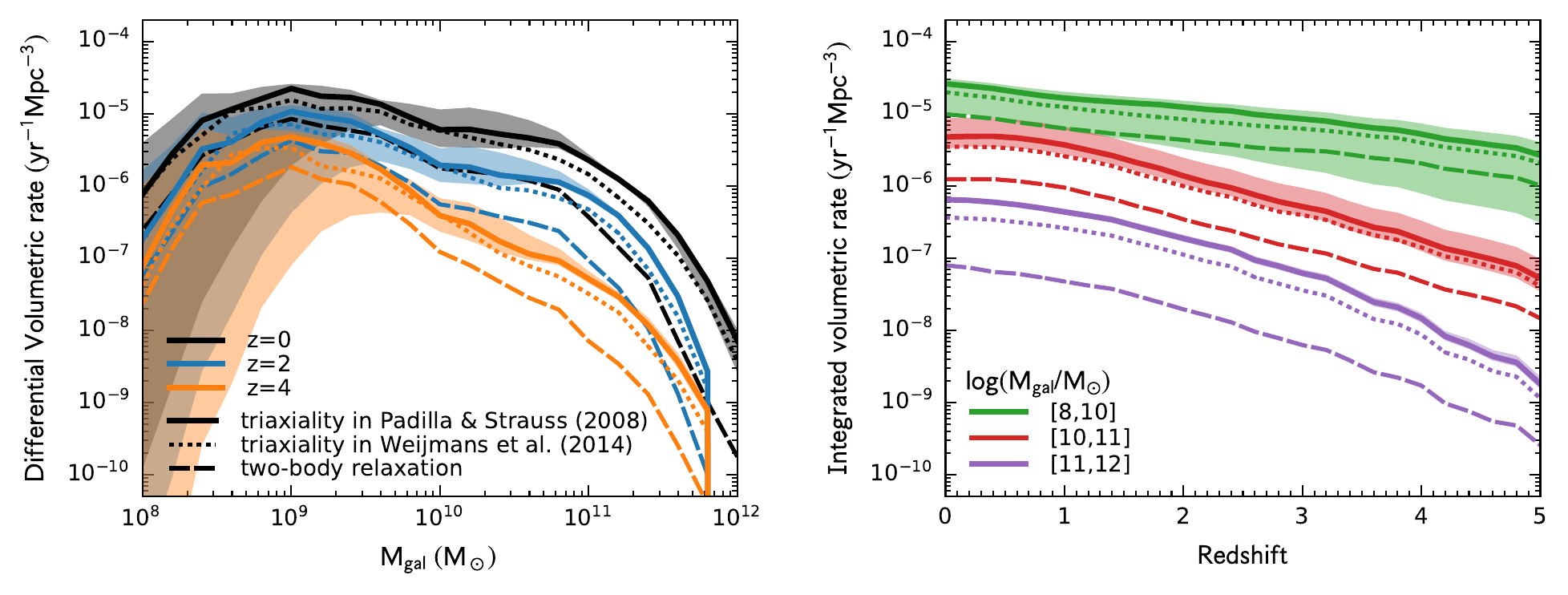}
\includegraphics[width=0.85\textwidth]{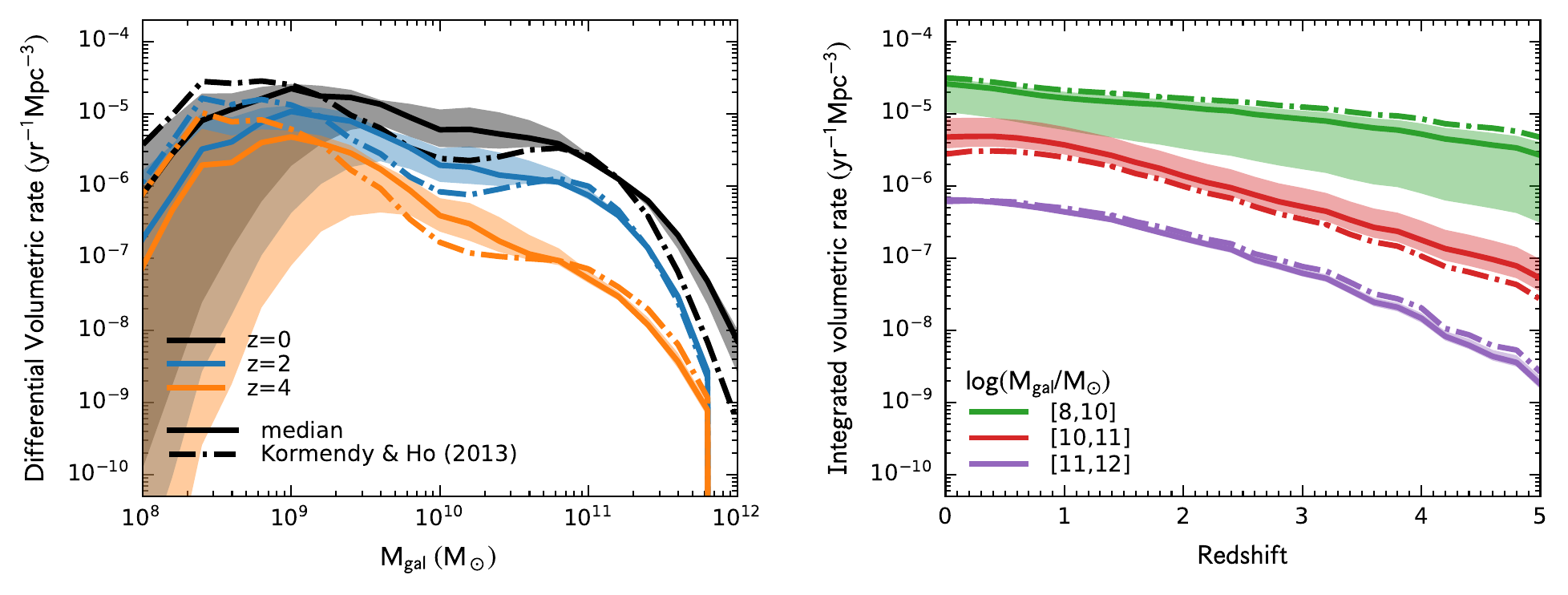}
\caption{Similar to Figure~\ref{fig:mock_rvol_Mgal} or
\ref{fig:mock_rvol_Mgal_m8}, but excluding galaxies with MBH mass $M\bh>
10^9\msun$. The volumetric stellar disruption rates in bright galaxies
($\sim 6\times 10^{-7}\pyr\pmpc$ at $z=0$ for $M\gal\ga 10^{11}\msun$) are
about 2\% of the stellar tidal disruption rates in all the
galaxies at $z=0$. } \label{fig:mock_rvol_Mgal_m9} \end{figure*}

In Figure~\ref{fig:mock_rvol_Mgal}, we show the differential and the integrated
volumetric stellar consumption rates as a function of galaxy mass $M\gal$.  As
seen from figure, the estimated rates in low-mass galaxies with $M\gal\la
10^9\msun$ are quite uncertain due to the scatters of the BH--host galaxy
relations. The median integrated rates at $z=0$ are $\sim 3\times
10^{-5}\pyr\pmpc$ over the mass range of $M\gal\sim 10^8$--$10^{10}\msun$,
$\sim 5\times 10^{-6}\pyr\pmpc$ over the mass range of $M\gal\sim
10^{10}$--$10^{11}\msun$, and $\sim 9\times 10^{-7}\pyr\pmpc$
over the mass range of $M\gal\sim 10^{11}$--$10^{12}\msun$ (see the solid curves
in the right panels).  The MBHs in Figure~\ref{fig:mock_rvol_Mgal} is over the
mass range of $M\bh=10^5$--$10^{10}\msun$, and the results include the rates of
stars being swallowed by high-mass MBHs. To see the volumetric stellar tidal
disruption rates,
Figures~\ref{fig:mock_rvol_Mgal_m8}--\ref{fig:mock_rvol_Mgal_m9} show the
results obtained with an MBH mass cut, i.e., $M\bh\le 10^8\msun$ (for
Schwarzschild BHs) and $M\bh\le 10^9\msun$ (for extremely spinning Kerr BHs),
respectively. In Figure~\ref{fig:mock_rvol_Mgal_m8}, the stellar tidal
disruption rates in bright galaxies ($\sim 3\times 10^{-7}\pyr\pmpc$ at $z=0$
for $M\gal\sim 10^{11}\msun$) are suppressed significantly, compared to the
stellar consumption rates shown in Figure~\ref{fig:mock_rvol_Mgal}.  
The suppresion is mainly due to the dearth of relatively low-mass central MBHs
(i.e., $M\bh\la 10^8\msun$) in these bright galaxies. The suppression trend is
also shown by \citet{Pfister20} (see the lower panel of Figure~4 therein).
However, Figure~\ref{fig:mock_rvol_Mgal_m9} shows that the volumetric stellar
disruption rates in bright galaxies ($\sim 6\times 10^{-7}\pyr\pmpc$ at $z=0$
for $M\gal\ga 10^{11}\msun$) are higher than those in
Figure~\ref{fig:mock_rvol_Mgal_m8} and up to 
$\sim$2\% of the stellar tidal disruption rates in all the galaxies at $z=0$.

\subsection{Stellar compact remnants}

One straightforward application of the above results is to roughly estimate the
rates of stellar compact remnants consumed by central MBHs, including white
dwarfs, neutron stars, and stellar-mass BHs. A crucial difference of the
consumption of compact remnants from the consumption of normal stars is that
the tidal radii of compact remnants $r\rmt$ are smaller than $r\rmswl=
4GM\bh/c^2$ (for Schwarzschild BHs) for a wide range of
$M\bh=10^5$--$10^{10}\msun$, and the compact remnants in the loss cone are
swallowed whole by central MBHs, with gravitational-wave bursts. 
The consumption rate of a given population of compact remnants is the product
of the number fraction of this population among normal stars and the stellar
consumption rate (obtained by setting $r\rmt=r\rmswl$).
Here we consider three compact remnant populations: white dwarfs (WDs), neutron
stars (NSs), and stellar-mass BHs (BHs).  If their number fractions among normal
stars are $f_{\rm cr}=$0.1, 0.01 and 0.001, respectively (e.g., see
\citealt{HA05}),
our calculations show that their integrated volumetric consumption rates over
the MBH mass ranges of $[10^5\msun,10^7\msun]$, $[10^7\msun,10^8\msun]$, and
$[10^8\msun,10^9\msun]$ are $f_{\rm cr}$ times of $1.3\times 10^{-5}\pyr\pmpc$,
$1.3\times 10^{-6}\pyr\pmpc$, and $2.0\times 10^{-6}\pyr\pmpc$, respectively,
where the MBH-bugle mass relation of \citet{Kormendy13} is used.  Note that the
above rates can be changed due to mass segregation at galactic centers and any
other mechanisms causing the number density distributions of compact remnants
to be different from those of normal stars (e.g., see \citealt{Alexander17}). 

\section{Conclusions}
\label{sec:concl}

We have investigated the rates of stellar consumption (either being tidally
disrupted or swallowed whole) by central MBHs and their distributions in the
realistic universe.  By adopting observational galaxy triaxiality shape
distributions and observational galaxy surface brightness distributions, we
study the effects of stellar orbital precession in triaxial galaxies to enhance
the number of stars that can move to the vicinity of the MBHs, as well as the
effects of two-body relaxation. By incorporating the refilling rates into the
loss cone and the draining rates of the loss regions in triaxial galaxies with
the MBH/galaxy demography, we obtain the following main results on the
distributions of both the per-galaxy and the volumetric stellar consumption
rates and their dependence on MBH and host galaxy properties.

The average (per-galaxy) stellar consumption rates in triaxial galaxies
$\mathscr{F}_{\rm consp}$ range from $\sim 5\times 10^{-4}\pyr\pgal$ to
$\sim 6\times
10^{-3}\pyr\pgal$ at $z=0$ and have a positive correlation with central MBH
masses for $M\bh\sim 10^7$--$10^9\msun$, and they range from $\sim 5\times
10^{-4} \pyr\pgal$ to $\sim 3\times 10^{-3} \pyr\pgal$ at $z=0$ and have a negative
correlation for $M\bh\sim 10^5$--$10^7\msun$. The rates at the peaks of the
stellar consumption rate distributions increase with increasing $M\bh$, being
around $10^{-4}\pyr\pgal$ at $z=0$ for $M\bh\sim 10^5$--$10^7\msun$ and ranging
from $\sim 3\times 10^{-4} \pyr\pgal$ to $\sim 3\times 10^{-3}\pyr\pgal$ at $z=0$ for
$M\bh\sim 10^8$--$10^9\msun$. The enhancement in the stellar consumption rates
caused by adding the effects of galaxy triaxial shapes is by a factor of
$\sim$3--5 for
$M\bh\la 10^7\msun$ and increases with increasing central MBH masses, which can
be up to one order of magnitude for MBHs with $M\bh\sim 10^8\msun$ and two
orders of magnitude for $M\bh\sim 10^9\msun$. The stellar consumption rates are
higher in the galaxies with steeper slopes in their surface brightness profiles
for $M\bh\la 10^7\msun$, and become insensitive to the inner slope of the
surface brightness profiles for higher MBH masses.

Regarding the differential volumetric consumption rate $\GtotVdiff$, the
increase caused by adding the effects of triaxial galaxy shapes is by a factor
of $\sim$3 for MBHs with mass $M\bh\la 10^7\msun$. For MBHs with larger masses,
the increase is larger. For example, $\GtotVdiff$ increases by a factor of
$\sim$5--8 at $M\bh\simeq 10^8\msun$ and by two orders of magnitude at $M\bh\simeq
10^9\msun$, compared to the rates obtained by considering only the loss-cone
refilling due to two-body relaxation in spherical gravitational potentials. The
TDE flaring rates are mainly dominated by the systems with low-mass MBHs, and
we have a median value of the integrated volumetric consumption rates ($\int
d\GVtotinte$) $\sim 3\times 10^{-5}\pyr\pmpc$ for all the MBHs with $M\bh\sim
10^5$--$10^8\msun$ at redshift $z=0$, and $\sim 1.4\times 10^{-6}\pyr\pmpc$ for
MBHs within $10^7$--$10^8\msun$.  The stellar swallowing rates are $\sim
8.0\times 10^{-7}\pyr\pmpc$ for MBHs with higher masses. Note that those
estimates are based on assuming the MBHs are Schwarzschild BHs. If the MBHs are
extremely spinning Kerr BHs (e.g., with the dimensionless spin parameter being
0.998), the upper mass limit of the MBH that can tidally disrupt a star with
solar mass and solar radius can increase from $\sim 10^8\msun$ to $\sim
10^9\msun$ (e.g., \citealt{MB20}), so that the TDE flare rates can be $\sim
10^{-6}\pyr\pmpc$ for MBHs with $M\bh\sim 10^8$--$10^9\msun$.

The volumetric stellar consumption rates decrease with increasing redshifts,
and the decrease is relatively mild for $M\bh\sim 10^5$--$10^7\msun$ and
stronger for higher $M\bh$.  Most of the stellar TDEs at $z=0$ occur at
galaxies with mass $M\gal\la 10^{11}\msun$. About 1\% of the TDEs can occur at
high-mass galaxies with $M\gal\ga 10^{11}\msun$ if all the MBHs are
Schwarzschild BHs, and up to $\sim$2\% of TDEs can occur at high-mass galaxies
with $M\gal\ga 10^{11}\msun$ if all the MBHs are extremely spinning Kerr BHs.

A statistically significant result of TDEs with an MBH mass spectrum and with
different galaxy properties (e.g., the inner slopes of surface brightness
profiles) obtained from observations will be helpful for a comparison with the
expectation from the work, e.g., to be achieved by future time-domain sky
surveys \citep{Stone20} by expanding the TDE sample by orders of magnitude and
with a sufficiently wide variety in galaxy properties.  Such a comparison would
provide significant constraints on the occupation number of MBHs in low-mass
galaxies, the relationship between MBH mass and host galaxy properties, and
general relativistic effects on TDEs, etc.

We thank the referee for perceptive comments.
This work was supported in part by the National Natural Science Foundation of
China under Nos.\ 11673001, 11273004, 10973001, 11690024, 11873056, and
11721303; the National Key R\&D Program of China (grant Nos.\ 2016YFA0400703,
2016YFA0400704); the Strategic Priority Program of the Chinese Academy of
Sciences (grant No.\ XDB 23040100); and National Supercomputer Center in
Guangzhou, China.

\end{document}